\documentclass{svjour3}

\usepackage{inputenc}

\usepackage[letterpaper,centering,margin=1in]{geometry}
\usepackage{natbib} 

\usepackage{amsfonts} 
\usepackage{amsmath}
\usepackage{amssymb}
\usepackage{mathtools}
\usepackage{bm}
\usepackage{url}

\usepackage[color=blue!30!white,textsize=tiny]{todonotes}
\usepackage{algorithm}
\usepackage[noend]{algpseudocode}

\usepackage{graphicx}

\DeclareMathOperator*{\argmin}{arg\,min}
\DeclareMathOperator*{\argmax}{arg\,max}
\DeclareMathOperator*{\diag}{diag}
\DeclareMathOperator*{\var}{Var}

\newtheorem{assumption}{Assumption}

\smartqed  

\newcommand{\edits}[1]{{\color{black}{#1}}}

\newcommand{\moreedits}[1]{{\color{black}{#1}}}

\DeclareUnicodeCharacter{2212}{\textendash}

\begin{document}

\title{Rate-optimal refinement strategies for local approximation MCMC
  \thanks{AS was supported by NSERC. AD and YM were supported by the SciDAC program of the DOE Office of Advanced Scientific Computing Research.}
}

\subtitle{}

\author{Andrew D. Davis \and Youssef Marzouk \and Aaron Smith \and Natesh Pillai}
\institute{A.~D.~Davis \at
                Courant Institute of Mathematical Sciences,
                New York, NY USA
              \email{davisad@alum.mit.edu}           
           \and
          Y.~M.~Marzouk \at
            Massachusetts Institute of Technology,
            Cambridge, MA USA
            \email{ymarz@mit.edu}
          \and
          A.~M.~Smith \at
            University of Ottawa,
            Ottawa, Canada
            \email{asmi28@uottawa.ca}
            \and
          N.~Pillai \at
            Harvard University,
            Cambridge, MA USA
            \email{pillai@fas.harvard.edu}
}

\titlerunning{Rate-optimal refinement for local approximation MCMC}        

\authorrunning{Davis et al.} 

\maketitle

\begin{abstract}
Many Bayesian inference problems involve target distributions whose density functions are computationally expensive to evaluate. Replacing the target density with a local approximation based on a small number of carefully chosen density evaluations can significantly reduce the computational expense of Markov chain Monte Carlo (MCMC) sampling. Moreover, continual refinement of the local approximation can guarantee asymptotically exact sampling. We devise a new strategy for balancing the decay rate of the bias due to the approximation with that of the MCMC variance. We prove that the error of the resulting local approximation MCMC (LA-MCMC) algorithm decays at roughly the expected $1/\sqrt{T}$ rate, and we demonstrate this rate numerically. We also introduce an algorithmic parameter that guarantees convergence given very weak tail bounds, significantly strengthening previous convergence results. Finally, we apply LA-MCMC to a computationally intensive Bayesian inverse problem arising in groundwater hydrology.

\keywords{Markov chain Monte Carlo \and local regression \and Bayesian inference \and surrogate models \and sampling methods}
\PACS{02.50.−r \and 02.50.Ng \and 02.50.Tt \and 02.70.Uu}
\end{abstract}

\section{Introduction}
\label{sec:intro}

Markov chain Monte Carlo (MCMC) is a sampling algorithm that {in principle} can simulate any probability distribution. {In practice,} many MCMC algorithms are infeasible when target density evaluations are computationally intensive. In this paper, we present an algorithm that mitigates this issue, achieving two desirable properties:  (i) vastly reducing the number of target density evaluations per MCMC step, while (ii) \textit{provably} retaining a convergence rate roughly equal to that of the original expensive algorithm. 

Challenges associated with computationally intractable target densities are extremely well known in the Monte Carlo literature. We now give a short survey of approaches to the problem, emphasizing the contributions of the current paper.
Replacing the intractable target density with an approximation or  ``surrogate model'' is the simplest way to significantly reduce the computational cost of MCMC. Replacing the target density with a \textit{fixed} surrogate, however, introduces a non-vanishing bias \citep{MarzoukXiu2009, Cotteretal2010, Bliznyuketal2012, LiMarzouk2014, Cui2016etal, stuart2018posterior}. 
The obvious solution to the problem of non-vanishing bias is to iteratively refine the surrogate during sampling. This strategy introduces some non-trivial technical difficulties: continual surrogate refinement within MCMC results in a process that is not Markovian. Such non-Markovian processes can have surprising and terrible convergence properties even when the surrogate has very low pointwise error \citep{LatuszynskiRosenthal2014}, sometimes failing to converge at all. In our previous work \citep{Conradetal2016, Conradetal2018}, we presented an MCMC algorithm with a continually refined surrogate based on \textit{local approximations}. We showed that this algorithm avoided the worst of these convergence problems and produced asymptotically exact results under certain strong assumptions. 

These strong assumptions are symptoms of a practical problem with how local approximation and MCMC interact in the ``tails'' of the target distribution. \citet{Conradetal2016, Conradetal2018} built a piecewise polynomial approximation by solving a local regression problem, using a small number of nearby exact target density evaluations. Refinements triggered by a cross-validation heuristic on the acceptance ratio added new points to the set of target density evaluations. Ideally, this approach should avoid expensive density evaluations in regions of low probability---particularly the tails of the distribution. In practice, there is an important tension: the cross-validation heuristic requires many \textit{more} tail evaluations than a good global design would suggest, but gives many \textit{fewer} tail evaluations than are sometimes required for global convergence and stability of the MCMC algorithm. 

In \citet{Conradetal2016, Conradetal2018} we showed that this tension could be avoided under certain strong tail conditions, with the additional requirement of randomly triggered surrogate refinements to ensure convergence. 

In more general settings, the underlying tension remains and leads to two serious issues which we address in the present paper.
First is the behavior of the cross-validation heuristic noted above: 
it demands more refinements in low probability regions than the goal of ``good'' approximation of the target density (in a natural sense that we will make precise later) would dictate. Second, and even worse, local polynomial approximations typically have very bad tail behavior, even failing to be integrable unless nearly every point in the tail has a density evaluation. 

To resolve the first issue, we introduce here a new surrogate refinement strategy that relaxes the acceptable error threshold in the tails of the distribution. The refinement strategy also \emph{balances} the rate of decay of the surrogate bias with that of Monte Carlo variance, using a local error indicator to characterize the bias. To resolve the second issue, we introduce a new correction term for the Metropolis-Hastings acceptance ratio that allows the algorithm to retain good convergence properties even under very weak assumptions about tail behavior, and without requiring integrability of the surrogate. 
The result is a local approximation MCMC (LA-MCMC) algorithm that tends to require far fewer density evaluations (resulting in faster runs) while delivering robust performance for a much broader class of target distributions. From a practical point of view, work such as \citet{Conradetal2016, Conradetal2018, Angelikopoulosetal2015} showed that the number of expensive target density evaluations can be reduced by orders of magnitude with minimal impact on the accuracy of target expectation estimates. Our new algorithm further reduces the number of density evaluations to nearly the rate-optimal number, without sacrificing MCMC stability.

From a theoretical point of view, we \textit{guarantee} that LA-MCMC converges quickly by providing useful bounds on the convergence of the algorithm after a \textit{finite} number of MCMC steps.  Our main result is that the mean-square error (MSE) of our LA-MCMC estimates decays at roughly the expected $1/T$ rate, where $T$ is the number of MCMC steps. We also show that LA-MCMC converges under very weak tail bounds that hold for many statistical examples. Comparatively, \citet{Conradetal2016} only established a law of large numbers that applied for special tail shapes. 

Using our optimal surrogate refinement strategy, the refinement rate (i.e., the rate at which new density evaluations are demanded) decreases with $T$. This implies that the decay of the MSE as the number of density evaluations $n$ increases is not only faster than the $1/n$ rate expected for standard geometrically ergodic MCMC, but also that this convergence rate (in $n$) may accelerate as the number of MCMC steps $T$ increases. 

\edits{
\paragraph{Other related work.}
There have been many attempts to accelerate MCMC for models that are expensive to compute. We have already discussed work based on the construction of surrogate models/approximations for the target density. This discussion was not exhaustive, and we point to the survey paper \citet{llorente2021survey} for a broader overview of various other techniques and when they are most useful. 
We also briefly describe several approaches that are quite different from ours.
The multi-level MCMC approach of \citet{Dodwelletal2015}, and related multi-index extensions \citep{jasra2018multi}, rely on predefined hierarchies of models---e.g., corresponding to different mesh refinements of an underlying differential equation model---that induce trade-offs between computational cost and accuracy. These approaches can drastically reduce the number of expensive model evaluations and total wallclock time of an MCMC run, but require a careful understanding of numerical approximation errors to achieve optimal convergence rates. Another part of the literature, including \citet{kaipio2007statistical,chkrebtii2016bayesian} and other related work, creates statistical models of numerical discretization error in ODE or PDE models. 
The delayed-acceptance MCMC method (e.g., \citet{ChristenFox2005, Cuietal2011}) can also exploit a hierarchy of models, but instead ``screens'' MCMC proposals through the approximate model(s) and evaluates the expensive target density at least once for each accepted sample, thus reducing cost by a constant factor. 

The MCMC method of \citet{doi:10.1080/10618600.2016.1231064} makes central use of delayed acceptance, but also builds surrogates through local approximation of the target density---and in that sense has many interesting links to the present approach. We summarize some of the similarities and differences as follows. First, rather than using a local polynomial approximation as we do here, \citet{doi:10.1080/10618600.2016.1231064} approximates the target density (or likelihood) with a simpler weighted average of its values at the $k$ nearest previous evaluation points. In principle, however, both algorithms can be run with a wide variety of local approximations. A more substantive distinction is that \citet{doi:10.1080/10618600.2016.1231064} fundamentally relies on the delayed-acceptance construction, and additionally on the presence of a standard non-adaptive Metropolis transition kernel employing the expensive target, to ensure ergodicity. 
The ratio of model evaluations to MCMC steps in the approximate MCMC algorithm is thus bounded below by a nonzero constant.
Moreover, to ensure diminishing adaptation, not all of the expensive target density evaluations are used to inform the surrogate, which reduces efficiency. 
In contrast, both \citet{Conradetal2016,Conradetal2018} and the new construction we propose here do not employ delayed acceptance, and instead carefully control the convergence of the local approximation to ensure ergodicity for the exact target.
All target density evaluations are used to build the surrogate, and we use a controlled, rather than opportunistic, design strategy to choose new evaluation points. 

A key improvement of the present work over both \citet{doi:10.1080/10618600.2016.1231064} and \citet{Conradetal2016,Conradetal2018} is that our new algorithm has much stronger theoretical convergence guarantees, and is applicable to a broader class of target distributions. For instance, earlier local approximation methods assume that the target distribution has tails that are very light or otherwise very special (\citet{doi:10.1080/10618600.2016.1231064} assume a uniform minorization condition; \citet{Conradetal2016} assumes either compact support or something very similar to a uniform minorization condition). Under such conditions, these efforts showed that their algorithms are ergodic (that is, Monte Carlo averages converge at some unspecified rate).
Our current effort substantially improves on all of these: we allow the ratio of model evaluations to MCMC steps to decay to zero quite quickly, we allow the target distribution to have quite general tails, and we give a nearly-optimal bound on the rate of convergence. We emphasize that these improvements are not merely theoretical; they represent real improvements in algorithm performance. For example, it is shockingly easy to write down an algorithm that looks quite a bit like the one in our present paper that \moreedits{\textit{does not}} converge at the correct rate or \textit{does} require the number of model evaluations to grow roughly linearly in the number of MCMC steps (see, e.g., Section~\ref{sec:example-banana}).

\paragraph{Scope.}
This paper focuses on the core problems of surrogate construction and refinement, tail correction, and convergence analysis, all in the context of a Metropolis--Hastings-type scheme. Many variations and extensions are possible, and indeed it is often fruitful to \emph{merge} the most important techniques presented in various papers. For instance, one could consider using gradient information from the surrogate within the Metropolis proposal, as in the MALA scheme of \citet{Conradetal2018}. \citet{doi:10.1080/10618600.2016.1231064} has an extensive discussion of $k$d-trees and their use in facilitating fast nearest neighbor searching; we use $k$d-trees in our implementation as well (see Section~\ref{sec:numerics}), but without some of the online point addition and re-balancing heuristics proposed in that paper. \citet{doi:10.1080/10618600.2016.1231064} also consider a pseudo-marginal version of their approximate MCMC algorithm, where only noisy unbiased evaluations of the expensive target density are available. An exhaustive evaluation and comparison of these methods is beyond the scope of the present work. A complete implementation of our algorithm is available within the software package MUQ (\url{https://muq.mit.edu}), so that users can combine our LA-MCMC scheme with more advanced transition kernels or other computational tools that improve performance. 

\moreedits{We also note that our emphasis here is on inference problems with expensive models (for instance, numerical discretizations of partial differential equations), but in \emph{moderate} parameter dimensions (for instance, $d=9$ in our last example, or $d=12$ in  \citet{Conradetal2018}). 
Posterior sampling in higher dimensional settings presents an additional set of challenges, outside of the present scope. Moreover, function approximation (and hence surrogate modeling) in high dimensions is generically subject to the curse of dimensionality, unless one can exploit some special structure.} For example, \citet{Cui2016etal} combines parameter-space dimension reduction with intrusive model reduction methods for expensive likelihoods. It should be feasible to combine our LA-MCMC approach with dimension reduction methods for Bayesian inverse problems \citep{cui2014likelihood,constantine2016accelerating,zahm2018certified} in a similar way.

}

\paragraph{Organization of the paper.} In Section~\ref{sec:numerics} we present our new algorithm, including the refinement scheme, the bias-variance tradeoff, and the tail correction. Section~\ref{sec:theory} describes our main theoretical results. Section~\ref{sec:examples} provides a range of numerical examples, beginning with simple configurations intended to illustrate specific features and variations of the algorithm, and culminating in the inference of a spatially distributed coefficient in a set of nonlinear PDEs.

\section{Local approximation MCMC}
\label{sec:numerics}

We assume that our target distribution has a density $\pi(x) = \exp{\mathcal{L}(x)}$ on $\mathbb{R}^d$. Our goal is to construct an algorithm that exploits regularity in $\pi$ to reduce the computational cost of simulating from the target distribution. As discussed in the Introduction, replacing target density evaluations with a continually and infinitely refined surrogate model, within MCMC, can asymptotically generate samples from the exact target distribution \citep{Conradetal2016}. Given a finite number of MCMC samples, however, replacing the target density with such an approximation introduces a \textit{surrogate bias}. In this section, we present a LA-MCMC algorithm that extends \citet{Conradetal2016} in two significant ways: (i) we use the trade-off between surrogate bias and Monte Carlo variance to develop a refinement strategy with near-optimal convergence guarantees; and (ii) we introduce a parameter that allows LA-MCMC to easily characterize distributions with heavier tails given little \textit{a priori} knowledge, while enabling additional speedups when the user has substantial \textit{a priori} knowledge.

Our approach will build a \textit{local polynomial} approximation \citep{Connetal2009,Kohler2002,Stone1977} of a function $g:\mathbb{R}^{d} \to \mathbb{R}$, where $g$ is chosen such that evaluating $\pi(x)$ is trivial given $g(x)$. For instance, $g$ might be the logarithm of the target density, $\mathcal{L}$; or it might be the log-likelihood function in a Bayesian setting, if the prior density is relatively simple to evaluate. Let the ``evaluated set'' $\mathcal{S}_n = \{x_1, \hdots, x_n\}$ comprise the set of parameter values $x_i \in \mathbb{R}^d$ at which we have evaluated $g(x_i)$. For now, we take this set as given; later, we discuss how to construct it. With an appropriate construction of $\mathcal{S}$, local approximations allow refinement to focus on regions where the target distribution has greater mass. Polynomial approximations are appealing in this setting because they can be built easily, are cheap to evaluate, and have known analytic derivatives.

\subsection{Local polynomial approximations}

We construct local polynomial approximations using weighted regression \citep{Kohler2002}. Let $\mathcal{P}$ be a polynomial space over $\mathbb{R}^{d}$. The approximation at each $x \in \mathbb{R}^{d}$ is $\hat{g}(x, \mathcal{S}_n)$ such that 
\begin{equation}
    \hat{g}(x, \mathcal{S}_n) = \argmin_{m \in \mathcal{P}}{\sum_{i=1}^{n} (m(x_i)-g(x_i))^2 W(x, x_i)},
    \label{eq:local-polynomial-estimate}
\end{equation}
where $W(x, x^{\prime})$ is a locally supported kernel. Typically, we choose the $k$-nearest neighbor kernel 
\begin{equation}
    W(x, x^{\prime}) = \begin{cases}
    1 & \mbox{if $x^{\prime} \in \mathcal{B}_k(x)$} \\
    0 & \mbox{otherwise,}
    \end{cases}
    \label{eq:hat-kernel}
\end{equation}
where $\mathcal{B}_k(x)$ is the smallest ball centered at $x$ containing $k$ elements of $\mathcal{S}_n$. Define a basis $\Phi = \{\phi_j\}_{j=1}^{q}$ for $\mathcal{P}$, let $\boldsymbol{\phi}(x) = [\phi_1(x), \hdots, \phi_q(x)]^T$, and define $\boldsymbol{x}_k(x, \mathcal{S}_n) = \{\tilde{x} \in \mathcal{S}_n: W(x, \tilde{x})>0\}$. The elements of $\boldsymbol{x}_k(x, \mathcal{S}_n)$ are the $k$ nearest neighbors to $x$ in $\mathcal{S}_n$. The optimal polynomial kernel estimate is 
\begin{equation}
    \hat{g}(x, \mathcal{S}_n) = \boldsymbol{\phi}(x)^T \mathbf{a}(x, \boldsymbol{x}_k(x, \mathcal{S}_n))
\end{equation}
such that 
\begin{equation}
    \mathbf{a}(x, \boldsymbol{x}_k) = \argmin_{\boldsymbol{\alpha} \in \mathbb{R}^{q}}{\|\mathbf{V}(\boldsymbol{x}_k) \boldsymbol{\alpha} - \mathbf{g}(\boldsymbol{x}_k)\|_{\mathbf{W}(x, \boldsymbol{x}_k)}^2},
    \label{eq:local-polynomial-estimate-matrix-form}
\end{equation}
where $\|\mathbf{x}\|_{\mathbf{A}}^2 = \mathbf{x}^T \mathbf{A} \mathbf{x}$, 
\begin{subequations}
\begin{equation}
    \mathbf{W}(x, \boldsymbol{x}_k) = \diag{[W(x, x_1), \hdots, W(x, x_k)]},    
\end{equation}
$\mathbf{V}(\boldsymbol{x}_k)$ is the Vandermonde matrix 
\begin{equation}
    \mathbf{V}(\boldsymbol{x}_k) = \left[ \begin{array}{c}
    \boldsymbol{\phi}(x_1)^T \\
    \vdots \\
    \boldsymbol{\phi}(x_k)^T
    \end{array} \right],
\end{equation}
and
\begin{equation}
    \mathbf{g}(\boldsymbol{x}_k) = \left[ \begin{array}{c}
     g(x_1) \\
     \vdots \\
     g(x_k)
     \end{array} \right].
\end{equation}
\end{subequations}
Assuming $\mathbf{V}(\boldsymbol{x}_k)$ has full column rank and $k \geq q$, the local polynomial approximation exists and is unique \citep{Kohler2002,Stone1977}. The solution to \eqref{eq:local-polynomial-estimate-matrix-form} is 
\begin{equation}
    \mathbf{a}(x, \boldsymbol{x}_k(x, \mathcal{S}_n)) = (\mathbf{V}^T \mathbf{W} \mathbf{V})^{-1} \mathbf{V}^T \mathbf{g}.
\end{equation}

\subsubsection{Error analysis}
\label{sec:erroranalysis}

Now we derive local error bounds that, for a fixed evaluated set $\mathcal{S}_n$, depend on the number of nearest neighbors, the size of the ball containing them, and the local behavior of the surrogate. We assume that the kernel $W(x, x^{\prime})$ is the hat kernel defined in \eqref{eq:hat-kernel} and that $\mathcal{P}$ is the space of polynomials of degree less than or equal to $p$; 
we write $q = \text{dim}( \mathcal{P})$. 

We first investigate the local behavior of the polynomials in the ball $\mathcal{B}_k(x)$. Let 
\begin{equation*}
\Delta(x) = \max_{x_i \in\boldsymbol{x}_k(x,\mathcal{S}_n)}{\|x-x_i\|}
\end{equation*}
be the radius of $\mathcal{B}_k(x)$---the ball that contains $k$ points in the evaluated set $\mathcal{S}_n$. We note that the radius of $\mathcal{B}_k(x)$ is not uniform in $x \in \mathbb{R}^{d}$. Analogous to \eqref{eq:local-polynomial-estimate-matrix-form}, define $k$ Lagrange polynomials
\begin{equation}
    \lambda_j(x) = \argmin_{m \in \mathcal{P}}{\sum_{i=1}^{k} (m(x_i)-\delta_{ij})^2},
\end{equation}
where $\delta_{ij}$ is the Dirac delta. Here, the summation is over the $k$ nearest neighbors to $x$, i.e., $x_i \in \boldsymbol{x}_k(x)$. Let $\boldsymbol{\lambda}(x) = [\lambda_1(x), \hdots, \lambda_k(x)]^T$. The nearest neighbors $\boldsymbol{x}_k(x)$ are called $\Lambda$-poised in $\mathcal{B}_k(x)$ if
\begin{equation}
    \max_{x^{\prime} \in \mathcal{B}_k(x)}{\|\boldsymbol{\lambda}(x^{\prime})\|_2} = \Lambda_2(x) \leq \Lambda.
    \label{eq:poisedness}
\end{equation}
This definition of $\Lambda$-poisedness is slightly different than in \citet{Connetal2009}, which uses the infinity norm. Since the number of nearest neighbors $k$ is fixed and finite, however, 
these definitions are equivalent. $\Lambda$-poisedness measures how well distributed the points are within the ball. For example, consider points $i$ and $j$ in a unit ball and quadratic polynomials. The $i^{\text{th}}$ Lagrange polynomial is close to one at point $i$ and close to zero at point $j$. (If the number of nearest neighbors $k$ is exactly the number required to interpolate ($k=q$) then the Lagrange polynomial will be exactly one or zero at these points.) 
The quadratic Lagrange polynomial is narrower if the points are close together and wider if they are far apart. The wider parabola has a smaller poisedness constant than the narrower one. 

 Assuming $\Lambda$-poisedness and that $g(x)$ is at least $(p+1)$ times differentiable, we have the error bound
\begin{align}
    \vert \hat{g}(x, \mathcal{S}_n) - g(x) \vert \leq \frac{k}{(p+1)!} \Lambda \Delta(x)^{p+1} \sup_{x^{\prime} \in \mathcal{B}_k(x)}{g^{(p+1)}(x^{\prime})},
    \label{eq:local-error-bound}
\end{align}
where $p$ is the polynomial degree and $g^{(p+1)}(x)$ is the $(p+1)^{\text{th}}$ derivative of $g(x)$ \citep{Connetal2009}. The radius $\Delta(x)$ decreases as points are added to $\mathcal{S}_n$ and thus the error bound decreases, assuming that points are chosen in a way that maintains $\Lambda$-poisedness.

\subsubsection{Local refinements}
\label{sec:localrefinements}
Given a surrogate $\hat{g}(x, \mathcal{S}_n)$, we can improve its accuracy in a neighborhood around $x$ by performing a \textit{local refinement}. A local refinement adds a new point $x^{*}$ to the evaluated set---i.e., $\mathcal{S}_{n+1} = \mathcal{S}_n \cup \{x^{*}\}$---and thus allows the local error bound \eqref{eq:local-error-bound} to decrease. Randomly choosing a nearby point tends to form clusters \citep{RoteTichy1996}, however, and therefore refining using a random point inside $\mathcal{B}_k(x)$ fails to maintain $\Lambda$-poisedness.

We instead choose the refinement location based on the poisedness constant $\Lambda_2(x)$ defined in \eqref{eq:poisedness}. Computing $\Lambda_2(x)$ by solving the optimization problem defined in \eqref{eq:poisedness} also defines the point 
\begin{equation}
    x_{\lambda}(x) = \argmin_{x^{\prime} \in \mathcal{B}_k(x)}{\|\boldsymbol{\lambda}(x^{\prime})\|_2}.
    \label{eq:poisedness-refinement-location}
\end{equation}
If $\boldsymbol{x}_k(x)$ is poorly poised in $\mathcal{B}_k(x)$ ($\Lambda_2(x) \gg 1$) then the new point $x_{\lambda}(x)$ will be relatively far from any clusters in $\boldsymbol{x}_k(x)$ and will tend to improve the poisedness of the updated set. Conversely, if $\boldsymbol{x}_k(x)$ is well poised in $\mathcal{B}_k(x)$ ($\Lambda_2(x) \approx 1$), then the updated set remains well poised. We therefore refine the surrogate by setting $\mathcal{S}_{n+1} = \mathcal{S}_n \cup \{x_{\lambda}(x)\}$.

Having discussed {how} to refine, we must also discuss \textit{when}, i.e., under what conditions, to refine. Our primary criterion will be derived from the local error bound \eqref{eq:local-error-bound}; this process is described in Section~\ref{sec:bias-variance-trade-off}. In the meantime, we discuss an additional but natural secondary criterion, which is to trigger a refinement if the poisedness constant exceeds a user-prescribed threshold $\bar{\Lambda}$ (i.e., if $\Lambda_2(x) > \bar{\Lambda}$ then set $\mathcal{S}_{n+1} = \mathcal{S}_n \cup \{x_{\lambda}(x)\}$). In practice, we find that adding $x_{\lambda}(x)$ to the evaluated set whenever a refinement is triggered tends to maintain poisedness automatically, and that the secondary criterion thus rarely triggers refinements. Computing the poisedness constant by solving \eqref{eq:poisedness} can also be computationally burdensome (although still significantly cheaper than an expensive density evaluation). We therefore often ``turn off'' this secondary refinement criterion by setting $\bar{\Lambda} = \infty$, and only compute the poisedness constant when refinements are triggered by the primary criterion. 

\subsection{Sampling methods using local approximations}

Each step of a typical MCMC algorithm consists of three stages: (i) propose a new state, (ii) compute the acceptance probability, and (iii) accept or reject the proposed state. LA-MCMC adds a fourth stage---possibly refine the surrogate model---and replaces the target density evaluations in stage (ii) exclusively with cheaper surrogate evaluations; see Algorithm \ref{alg:la-mcmc}. The refinement frequency ensures that the error incurred by using a surrogate model balances the Monte Carlo variance.

\subsubsection{Bias-variance trade-off} \label{sec:bias-variance-trade-off}

We now present a heuristic for balancing bias and variance in LA-MCMC,  deferring a rigorous discussion to Section \ref{sec:theory}.

Fix a function $f$ and Markov chain $\{X_{t}\}_{t \geq 0}$ with stationary measure $\pi$. For $T \in \mathbb{N}$, denote by $\hat{\pi}_{T}(f) = \frac{1}{T} \sum_{t=1}^{T} f(X_{t})$ the usual Monte Carlo estimate of $\pi(f) = \mathbb{E}[f(X)]$---the expectation with respect to $\pi$. 
Under modest conditions (see, e.g.,  \citet{meyn2012markov}), the asymptotic bias exists and satisfies 
\begin{equation}
\lim_{T \rightarrow \infty} T \, \left \vert \mathbb{E}[\pi_{T}(f)] - \pi(f) \right \vert = 0
\end{equation}
while the asymptotic variance exists and satisfies
\begin{equation}
\lim_{T \rightarrow \infty} T \, \var{(\hat{\pi}_{T}(f))}  = C_v \in (0,\infty).
\end{equation}
When this happens, there is (asymptotically) no trade-off: the bias quickly becomes a negligible source of error.

In the context of LA-MCMC,  more care is required because of an additional bias from the surrogate model. We assume that $\pi$ has density $\pi(x) = \exp{ \mathcal{L}(x) }$, and view LA-MCMC as an ``approximate'' MCMC algorithm that tries to target\footnote{We note that our $\widehat{\pi}(x) = \exp \widehat{\mathcal{L}}(x)$ often fails to be a probability density. We resolve this technical problem in Section \ref{sec:theory}, and somewhat surprisingly our approach means that this problem has very little impact on the following heuristic calculations.} $\hat{\pi}(x) = \exp \widehat{\mathcal{L}}(x)$, where $\widehat{\mathcal{L}}(x) \approx \mathcal{L}(x)$. In particular, we assume that the surrogate $\widehat{\mathcal{L}}$ adheres to the error bound in \eqref{eq:local-error-bound}, which is a pointwise condition of the form: 
\begin{equation}
 \vert \widehat{\mathcal{L}}(x) - {\mathcal{L}}(x) \vert \leq C_{b} \Delta(x)^{p+1}
 \label{eq:transition-kernel-radius-bound}
\end{equation} 
for $C_b>0$. We will show in Section \ref{sec:theory} that the bias incurred from using this surrogate model inherits a similar error bound,
\begin{equation}
    \left| \hat{\pi}(f) - \pi(f) \right| \leq C_{b}^{\prime} \bar{\Delta}^{p+1},
    \label{eq:expectation-error}
\end{equation}
where $C_b^{\prime}>0$ and, informally, we can think of $\bar{\Delta}^{p+1}$ as the maximum of $\Delta(x)^{p+1}/V(x)$, where $V(x) > 0$ is a penalty function to be described below. We will thus use the local radius $\Delta(x)^{p+1}$ as an \textit{error indicator}, and sequentially refine our approximation so that $\bar{\Delta} = \bar{\Delta}(T)$ decays with time $T$.

We would like to balance the additional error introduced by the surrogate model against the usual ``Monte Carlo'' error of the original Markov chain. 
We choose our parameters so that the upper bound on the surrogate error decays at the same rate as the Monte Carlo error. 
To ensure that the pointwise surrogate error is small enough, we partition the chain into 
consecutive intervals called \emph{levels}, $\ell = 0, 1, 2, \ldots$, and prescribe a piecewise constant error threshold for each level $\ell$,
\begin{equation}
    \gamma_\ell(x) = \gamma_0 \ell^{-\gamma_1} V(x),
    \label{eq:error-threshold}
\end{equation}
with $\gamma_0, \gamma_1 > 0$.
Here, $V(x) > 0$ is a $\pi$-integrable penalty function that allows the error threshold to be larger in low-probability regions (e.g., in the tails of the distribution). We assume that $V(x)$ satisfies the Lyapunov inequality \eqref{eq:lyapunov}; this requirement guides how to choose the function. As a reasonable default, we set $V(x) = \exp{(\nu_0 \|x-\bar{x}\|^{\nu_1})}$, where $\nu_0>0$, $0 < \nu_1 \leq 1$, and $\bar{x}$ is a user-prescribed estimate of the centroid of the distribution (e.g., the mean or mode). During MCMC, we locally refine the surrogate model if the error indicator $\Delta(x)^{p+1}$ at the current state exceeds the current threshold $\gamma_\ell(x)$. 

Level $\ell$ ends at step $T_{\ell}$, when the MCMC variance is of the same size as the squared surrogate bias:
\begin{equation}
    \left ( \gamma_0 \ell^{-\gamma_1} \right )^2 \sim C_{v} T^{-1}_\ell.
    \label{eq:bias-variance}
\end{equation}
The number of steps in the $\ell^{\text{th}}$ level is, therefore, 
\begin{equation}
    T_{\ell} = \tau_0 \ell^{2\gamma_1},
\end{equation}
where $\tau_0 = C_v / \gamma_0^2$ is the length of the first level. In practice, we set the the level as a function of the MCMC step $t$, 
\begin{equation}
    \ell(t) = \lfloor (t/\tau_0)^{1/(2\gamma_1)} \rfloor.
    \label{eq:leveldefn}
\end{equation}
The length of each level, $\tau_{\ell} = T_{\ell}-T_{\ell-1}$, will then decrease with $\ell$ if $\gamma_1 < 0.5$. This is undesirable because then, beyond some critical step $t^* < \infty$, the level length will become less than one. If this occurs, we will need to increment the level more than once per MCMC step. Rather than addressing this complication, we simply require that lengths of the levels strictly increase with $\ell$ by setting $\gamma_1 > 0.5$. 

The constant $\tau_0$ is rarely known, and we treat it as an algorithmic parameter. Given the initial error threshold $\gamma_0>0$, error decay rate $\gamma_1>0.5$, initial level length $\tau_0 \geq 1$, and the Lyapunov function $V(x)$, we locally refine whenever the error indicator at an accepted state in the chain exceeds the decaying error threshold. See Algorithm \ref{alg:la-mcmc} for a complete summary. 

\subsubsection{Tail correction}
\label{sec:tailcorrection}

There is a major technical difficulty in making the heuristic from Section \ref{sec:bias-variance-trade-off} precise: it is possible that our surrogate function has small pointwise error as in \eqref{eq:transition-kernel-radius-bound}, but fails to be globally integrable (and thus cannot possibly be an unnormalized probability density). This results in large practical and theoretical difficulties: the associated stochastic process may wander off to infinity, and the pointwise bound in \eqref{eq:transition-kernel-radius-bound} will not give any bound on the Monte Carlo error. In previous work (e.g., \citet{Conradetal2016}), we avoided this problem by working on compact state spaces or obtaining much stronger pointwise approximations than \eqref{eq:transition-kernel-radius-bound}. Here, we obtain much stronger results by slightly tweaking our algorithm's acceptance probability.

Our tweak uses the Lyapunov function of the MCMC algorithm with exact evaluations, $V(x)$, to change the acceptance probability. Rather than computing the typical Metropolis-Hastings acceptance probability targeting $\hat{\pi}$, we slightly \textit{increase} the chances of moves that would decrease the Lyapunov function and \textit{decrease} the chances of moves that would increase it. \citet{roberts1996geometric} give very general conditions under which we can easily define the Lyapunov function.

More precisely, given the current state $x_t$ and proposed state $x^{\prime}_t$, we tweak the usual Metropolis-Hastings acceptance probability with the estimate 
\begin{equation}
    \log{\pi(x^{\prime}_t)} \approx \widetilde{\mathcal{L}}(x^{\prime}_t) \equiv \widehat{\mathcal{L}}(x^{\prime}_t) + Q_V(x_t, x^{\prime}_t),
    \label{eq:Lypunov-correction}
\end{equation}
where 
\begin{equation}
    Q_V(x_t, x^{\prime}_t) = \begin{cases}
    \ \, \, \eta (\gamma(x^{\prime}_t) + \gamma(x_t))   & \mbox{if } V(x'_t) < V(x_t)\\
    - \eta(\gamma(x^{\prime}_t) + \gamma(x_t)) \ & \mbox{if } V(x'_t) \geq V(x_t),
    \end{cases}
    \label{EqQvDef}
\end{equation} 
and $\gamma(\cdot) = \gamma_{\ell(t)}(\cdot)$ is defined by \eqref{eq:error-threshold} and \eqref{eq:leveldefn}. Here, $\widehat{\mathcal{L}}(x^{\prime}_t)$ is an approximation of the log-density using the local polynomial approximation. The factor $\gamma(x^{\prime}_t) + \gamma(x_t)$ causes the correction to become less apparent as the error threshold decays, 
$\gamma_{\ell(t)}(\cdot) \to 0$,
and $\eta \geq 0$ is a user-defined parameter to control the correction. 

\begin{algorithm}
 \begin{algorithmic}
    \State Set initial state $X_0$ and surrogate model $\widehat{\mathcal{L}}_0$ for the log-target density 
    \For{$t \gets 1$ to $\infty$}
    \State Propose $X^{\prime} \sim q_t(\cdot \vert X_t)$
    \State
    \State Possibly refine at $X_t$: $(\gamma(X_t), \widehat{\mathcal{L}}_t) = $ \Call{CheckAndRefine}{$X_t$, $\widehat{\mathcal{L}}_{t-1}$, $t$}
    \State
    \State Compute error threshold at the proposed point $$\gamma(X^{\prime}) = \gamma_0 (\lfloor (t/\tau_0)^{1/(2\gamma_1)} \rfloor)^{-\gamma_1} V(X^{\prime})$$
    \State
    \State Compute $\widetilde{\mathcal{L}}_t(X^{\prime})$ via \eqref{eq:Lypunov-correction} and evaluate the acceptance probability
    \begin{equation*}
        \alpha(X_t, X^{\prime}) = \min{\left(1, \exp{(\widetilde{\mathcal{L}}_t(X^{\prime})-\widehat{\mathcal{L}}_t(X_t))} \frac{q_t(X_t \vert X^{\prime})}{q_t(X^{\prime} \vert X_t)} \right)}
    \end{equation*}
    \State
    \State Accept/reject step:
    \begin{equation*}
        X_{t+1} = \begin{cases}
        X^{\prime} & \mbox{with probability } \alpha(X_t, X^{\prime}) \\
        X_t & \mbox{else}
    \end{cases}
    \end{equation*}
    \EndFor
    \Procedure{CheckAndRefine}{$x$, $\widehat{\mathcal{L}}$, $t$}
  \State Compute the local error threshold $$\gamma(x) = \gamma_0 (\lfloor (t/\tau_0)^{1/(2\gamma_1)} \rfloor)^{-\gamma_1} V(x)$$ and (optionally) compute the poisedness constant $\Lambda_2(x)$
  \State
  \If{$\Delta(x)^{p+1}>\gamma(x)$ or (optionally) $\Lambda_2(x) > \bar{\Lambda}$} 
    \State $\widehat{\mathcal{L}}^{*}$ = \Call{RefineSurrogate}{$x$, $\widehat{\mathcal{L}}$}
  \Else{}
    $\widehat{\mathcal{L}}^{*} = \widehat{\mathcal{L}}$
  \EndIf
  \Return The error threshold $\gamma(x)$ and $\widehat{\mathcal{L}}^{*}$
  \EndProcedure
  \State
  \Procedure{RefineSurrogate}{$x$, $\widehat{\mathcal{L}}$}
  \State Compute poisedness-based refinement location $x_{\lambda}(x)$ defined in \eqref{eq:poisedness-refinement-location}
  \If{$x_{\lambda}(x) \notin \mathcal{S}_n$} 
    Set $\mathcal{S}_n \gets \mathcal{S}_n \cup \{x_{\lambda}(x)\}$ and $n \gets n+1$
  \EndIf
  \If{$x_{\lambda}(x) \in \mathcal{S}_n$} 
    Randomly choose $X^{*} \in \mathcal{B}_k(x)$ and set $\mathcal{S}_n \gets \mathcal{S}_n \cup \{X^{*}\}$ and $n \gets n+1$
  \EndIf
  \Return The updated surrogate model $\widehat{\mathcal{L}}^{*}$ using $\mathcal{S}_n$
  \EndProcedure
 \end{algorithmic}
 \caption{Pseudocode for the LA-MCMC algorithm. LA-MCMC requires eight parameters: (i) the initial error threshold $\gamma_0>0$, (ii) the error threshold decay rate $\gamma_1>0.5$, (iii) the maximum poisedness constant $\bar{\Lambda}$, (iv) the length of the first level $\tau_0 \geq 1$, (v) the tail-correction parameter $\eta \geq 0$, (vi) the number of nearest neighbors $k$ used to construct the local polynomial surrogate, (vii) the degree of the local polynomial surrogate $p$, and (viii) a guessed Lyapunov function $V$.}
 \label{alg:la-mcmc}
\end{algorithm}

\edits{
\subsubsection{Algorithm parameters}

We briefly describe some heuristics for choosing the input parameters to Algorithm~\ref{alg:la-mcmc}. Ideally, we choose the initial threshold $\gamma_{0}$ so that the required ball size $\Delta$ on the first error level is similar to the standard deviation of the posterior. Although the latter quantity is unknown, we can often estimate it using derivative information around the posterior mode. We have found that the decay rate $\gamma_{1} > 0.5$ does not have much impact on the performance of the algorithm (see Section~\ref{sec:1d-example}). We normally set $\tau_0=1$. Typically we use quadratic surrogates, i.e., $p=2$; we further discuss this choice in Sections~\ref{sec:1d-example} and \ref{sec:tracer}. We usually prescribe the number of nearest neighbors to be $k = 2q$, where $q = \text{dim}( \mathcal{P})$ is the dimension of the polynomial space. (In principle we only need $k \geq q$, but the extra regression points add stability.)

There is usually a great deal of flexibility in the choice of Lyapunov function, as it should only have a large impact on the tails of the target distribution. In order to prevent the stochastic process from wandering to infinity, it must satisfy $\pi(V) < \infty$.  If one has knowledge of the tail behavior of the target density $\pi$, choosing $V(x) \propto 1/\sqrt{\pi(x)}$ often works well. If we merely have bounds on the tails, we denote by $\Pi$ a distribution with heavier tails than $\pi$ and try $V(x) \propto 1/\sqrt{\Pi(x)}$. We choose the tail correction parameter $\eta$ to be as small as possible without letting the chain wander out to infinity. Typically, setting $\eta=0$ and slowly increasing the parameter until the chains stop wandering is sufficient. Finally, the bound on the poisedness constant, $\bar{\Lambda}$, is mostly included for reasons related to our proof technique. As discussed in Section~\ref{sec:localrefinements}, we often set  $\bar{\Lambda} = \infty$, and even the theory allows for very large values.}

\edits{Efficiently finding $\boldsymbol{x}_k(x, \mathcal{S}_n)$, i.e., the $k$ nearest neighbors to a point $x$, is a non-trivial but fortunately well studied problem. We store points of the evaluated set $\mathcal{S}_n$ in $k$d-trees, which enable efficient nearest neighbor searching. Specifically, we use the $k$d-tree implementation in the library \texttt{nanoflann} \citep{nanoflann}, which is quite efficient; as reported in \citet{nanoflann}, it can build $k$d-trees from point sets of size $n=10^6$ in microseconds. (This benchmark is for $d=3$, but tree construction cost scales only linearly with $d$.) For the examples we have studied, which typically have $n < 10^4$ and $d \leq 12$, $k$d-tree construction cost is negligible. 
Our current implementation rebuilds the tree structure at each refinement step; this process ensures that the time needed to search the tree, a task invoked much more frequently, remains small. For problems where tree construction time is non-negligible, online point addition and re-balancing approaches, as discussed in \citet{doi:10.1080/10618600.2016.1231064}, could be beneficial to incorporate into the present workflow.

}

\section{Theoretical results} \label{sec:theory}

The main theoretical results of this paper are that the LA-MCMC algorithm described in Section~\ref{sec:numerics} inherits approximately the same $1/T$ convergence rate for the mean squared error as MCMC with exact evaluations and that our slight tweak to the acceptance probably allows LA-MCMC to converge even for heavier tailed distributions. We summarize the results and their implications here and refer to Appendix \ref{app:theory} for detailed discussion.

Denote by $K$ the transition kernel of a discrete-time Markov chain on $\mathbb{R}^{d}$ with unique stationary distribution $\pi_{X}$ that has density $\pi(x)$. We wish to study the stochastic processes that approximate $K$, despite not being Markov chains. Note that most of the calculations in this section are similar to those in the ``approximate'' Markov chain literature (see, e.g.,  \citet{Johndrowetal2015,Medinaetal2018,PillaiSmith2014,Rudolfetal2018}).  The biggest differences are:
\begin{enumerate}
    \item Our processes are not quite Markov chains. This does not substantially change any calculations, but does require us to be slightly more careful in a few steps.
    \item We make slightly atypical assumptions about our Lyapunov functions. These are justified in Section \ref{SubsecInheritLyap}.
\end{enumerate}

The purpose of the ``tweak'' in \eqref{eq:Lypunov-correction} was to force our main algorithm to satisfy the same Lyapunov condition as the baseline MCMC algorithm that it is approximating. To be more precise, we assume that the Metropolis-Hastings transition kernel $K$ associated with proposal kernel $q_t$ and target distribution $\pi_{X}$ satisfies:
\begin{assumption} (Lyapunov inequality). There exists $V: \mathbb{R}^{d} \to [1, \infty)$ and constants $0 < \alpha \leq 1$ and $0 \leq \beta < \infty$ so that 
\begin{equation}
\label{eq:lyapunov}
    (K V)(x, \cdot) \leq (1-\alpha) V(x) + \beta
\end{equation}
for all $x \in \mathbb{R}^{d}$. 
\label{assumption:lyapunov-inequality-simple}
\end{assumption}
\begin{assumption}
(Geometric ergodicity). Let Assumption \ref{assumption:lyapunov-inequality-simple} (or Assumption~\ref{assumption:lyapunov-inequality}; see Appendix \ref{app:theory}) hold. There exist $0 < R < \infty$ and $0 \leq \gamma < 1$ so that 
\begin{equation}
    \sup_{x \, : \, V(x) \leq 4 \beta / \alpha}{\|K^{s}(x, \cdot) - \pi(\cdot)\|_{TV}} \leq R \gamma^s
\end{equation}
for all $s \geq 0$.
\label{assumption:geometric-ergodicity}
\end{assumption}
Note that we denote by $K^{s}(x,\cdot)$ the $s$-step transition density for a chain sampled from $K$ with starting point $x$, so that $K^{1}(x,\cdot) = K(x,\cdot)$. 

We now link to Algorithm \ref{alg:la-mcmc} with the same proposal $q_t$ and target $\pi_X$. Let $\{ X_{t}\}_{t \geq 0}$ be the result of a run of this algorithm, 
%
and let $\{ \widehat{\mathcal{L}}_{t}\}_{t \geq 0}$ and $\{ X_{t}' \}_{t \geq 0}$ be the sequence of surrogates and proposals computed during the run. We make the following assumption about the ``goodness" of our approximation:

\begin{assumption} 
(Approximation goodness). The parameters of Algorithm \ref{alg:la-mcmc}  are such that the approximations $\widehat{\mathcal{L}}_{t}$ to the log-density satisfy
\begin{equation}
    \left \vert \widehat{\mathcal{L}}_{t}(X_{t}) - \mathcal{L}(X_{t}) \right \vert + \left \vert \widehat{\mathcal{L}}_{t}(X_{t}') - \mathcal{L}(X_{t}') \right \vert  \leq  \left \vert Q_{V}(X_{t}, X_{t}') \right \vert 
\end{equation}
deterministically, where $Q_{V}(X_{t}, X_{t}')$ is as defined in \eqref{EqQvDef}. 
\label{assumption:good-appr}
\end{assumption}
Fix a function $f : \mathbb{R}^{d} \to [-1,1]$ with $\pi(f) = 0$. Our main theoretical result is:
\begin{theorem}
Let Assumptions \ref{assumption:lyapunov-inequality-simple}, \ref{assumption:geometric-ergodicity}, and \ref{assumption:good-appr} hold and let $\{X_t\}_{t=1}^{T}$ be generated by Algorithm \ref{alg:la-mcmc} with starting point $x$. Then there exists a constant $C > 0$ so that
\begin{equation}
\left|\mathbb{E}\left[ \left( \frac{1}{T} \sum_{t=1}^{T} f(X_t) \right)^2 \right] \right| \leq C \frac{\log(T)^{3}}{T}
\end{equation}
for all $T > T_{0}(x)$ sufficiently large.
\label{thm:convergence-rate}
\end{theorem}
To give Theorem \ref{thm:convergence-rate} some context, the ``usual'' MCMC estimate satisfies $\mathbb{E}[(\hat{\pi}_{T}(f))^2] \leq C/T$ for all $T > T_{0}$ sufficiently large 
under the same Lyapunov and minorization assumptions (Assumptions \ref{assumption:lyapunov-inequality-simple} and \ref{assumption:geometric-ergodicity}---see \citet{meyn2012markov}). Furthermore, the usual bound is sharp. Our bound is nearly identical, giving up only logarithmic terms. Informally, this suggests that using our algorithm has nearly the same finite-time guarantees as typical MCMC algorithms and, depending on the specific problem, may be substantially cheaper to run.

\section{Numerical examples}
\label{sec:examples}
We present three numerical experiments. The first focuses on understanding the convergence of the LA-MCMC algorithm and the impact of various algorithmic parameters controlling the approximation. The second experiment illustrates the tail correction approach described in Section~\ref{sec:tailcorrection}. The third then demonstrates the practical performance of LA-MCMC in a computationally challenging large-scale application: an inverse problem arising in groundwater hydrology.

\subsection{One-dimensional toy example} \label{sec:1d-example}

\begin{figure}[h!]
\centering
    \includegraphics[width=0.475\textwidth]{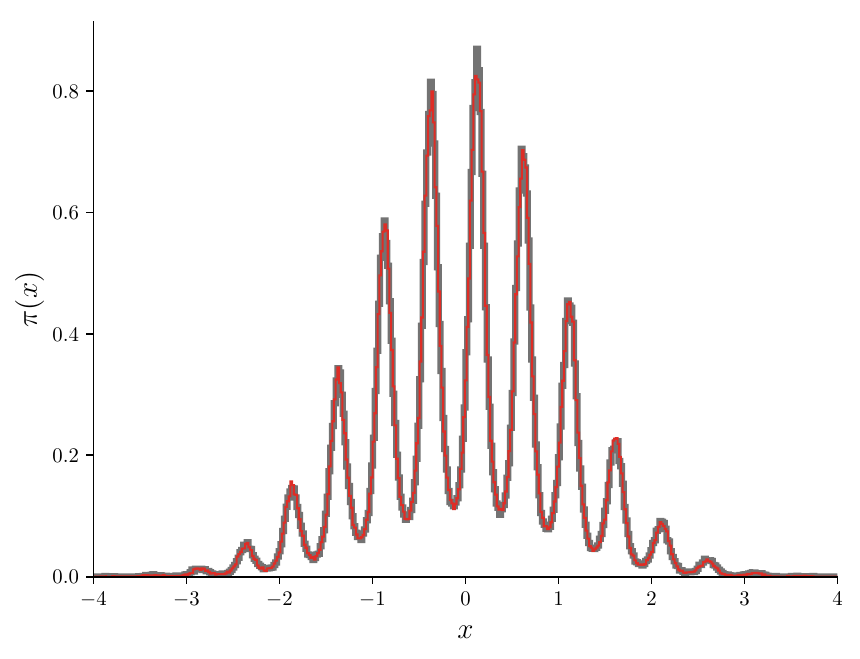}
    \caption{Binned MCMC samples computed with exact evaluations (grey line) and using local approximations (red line).}
    \label{fig:1d-example-density}
\end{figure}

\begin{figure}[h!]
\centering
    \includegraphics[width=0.475\textwidth]{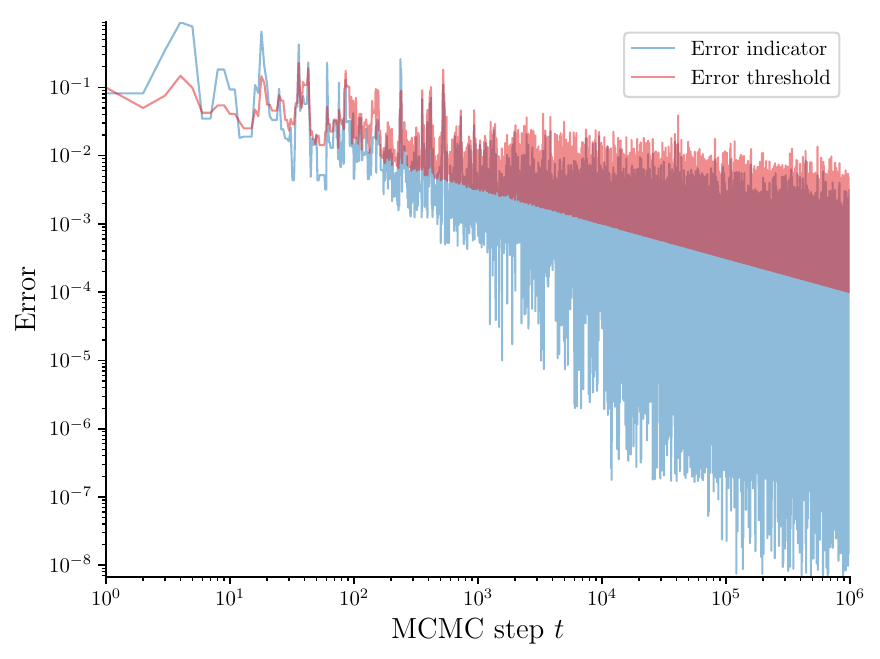}
    \caption{The local error indicator $\Delta(x)^{p+1}$ (blue line) and error threshold $\gamma_\ell(x)$ (Equation \eqref{eq:error-threshold}---red line) used to trigger refinement. The sharp lower bound for the error threshold (bottom border of the red region) is the error threshold with $V(x) = 1$. We, however, allow $V(x) = \exp(\|x\|)$ to be larger in the tails, which relaxes the error threshold in low probability regions. This example run triggered $472$ refinements. Many more (unnecessary) refinements would be required if the Lyapunov function did not relax the allowable error in the tails.}
    \label{fig:1d-example-error}
\end{figure}

We first use the one-dimensional density
\begin{equation}
    \log{\pi(x)} \propto -0.5 x^2 + \sin{(4 \pi x)}, \ x \in \mathbb{R}
\end{equation}
to demonstrate how changing various algorithmic parameters affects the performance of LA-MCMC. Figure \ref{fig:1d-example-density} shows binned MCMC samples computed with a random-walk Metropolis algorithm that uses exact target evaluations, compared with binned samples from an LA-MCMC algorithm that uses the same proposal. The sample histograms match very closely. Figure \ref{fig:1d-example-error} shows the error indicator $\Delta(x)^{p+1}$ and error threshold \eqref{eq:error-threshold} computed in a single run of LA-MCMC. Both the error indicator and threshold depend on the current state $X_t$: the indicator depends on the local ball size, and the threshold depends on the Lyapunov function $V(X_t) = \exp{(\|X_t\|)}$. Intuitively, the Lyapunov function relaxes the refinement threshold in the tails of the distribution and thus prevents excessive refinement in low probability regions, where the ball size $\Delta(x)$ tends to be large. 

LA-MCMC algorithmic parameters for the runs described here and below are $\gamma_0=0.1$,  $\bar{\Lambda} = \infty$, $\tau_0=1$, $\eta = 0$, $k=2(p+1)$, and $V(x) = \exp{(\|x\|)}$ (see Algorithm \ref{alg:la-mcmc}). We use local polynomial degree $p=2$ and $\gamma_1 = 1$ unless otherwise indicated.

Let $\sigma^2_t$ and $\breve{\sigma}^2_t$ be running $t$-sample estimates of the target variance computed using sequences of samples $\{X_t\}_{t>0}$ generated by MCMC with exact evaluations (which we refer to as ``exact MCMC'' for shorthand) and LA-MCMC, respectively. As a baseline for comparison, we also compute a `high fidelity' approximation of the variance from 50 realizations of exact MCMC: $\bar{\sigma}^2 = \sum_{i=1}^{50} \sigma_{T,i}^2$, with $T = 10^{6}$. Then we evaluate the error in variance estimates produced by exact MCMC and LA-MCMC,
\begin{equation}
    \begin{array}{ccc}
        e_t = \vert \bar{\sigma}^2 - \sigma_t^2 \vert & \mbox{and} & \breve{e}_t = \vert \bar{\sigma}^2 - \breve{\sigma}_t^2 \vert.
    \end{array}
\end{equation}
We compute the expectations of these errors by averaging $e_t$ and $\breve{e}_t$ over multiple independent realizations of each chain.  

The bias-variance trade-off used to construct our algorithm (see Section \ref{sec:bias-variance-trade-off}) ensures that the error in an expectation computed with LA-MCMC decays at essentially the same rate as the error in an expectation computed with exact MCMC. However, we need to tune the initial error threshold $\gamma_0$ and initial level length $\tau_0$ to ensure that the expected errors are of the same magnitude. In general, we set $\tau_0=1$. The initial error threshold also determines the initial 
local radius $\Delta(x) = \gamma_0^{1/(p+1)} V(x)$. As a heuristic, we choose $\gamma_0$ so that the initial radius is smaller than the radius of a ball containing the non-trivial support of the target density. 

\begin{figure}
  \centering
  \begin{tabular}{@{}c@{}}
    \includegraphics[width=0.475\textwidth]{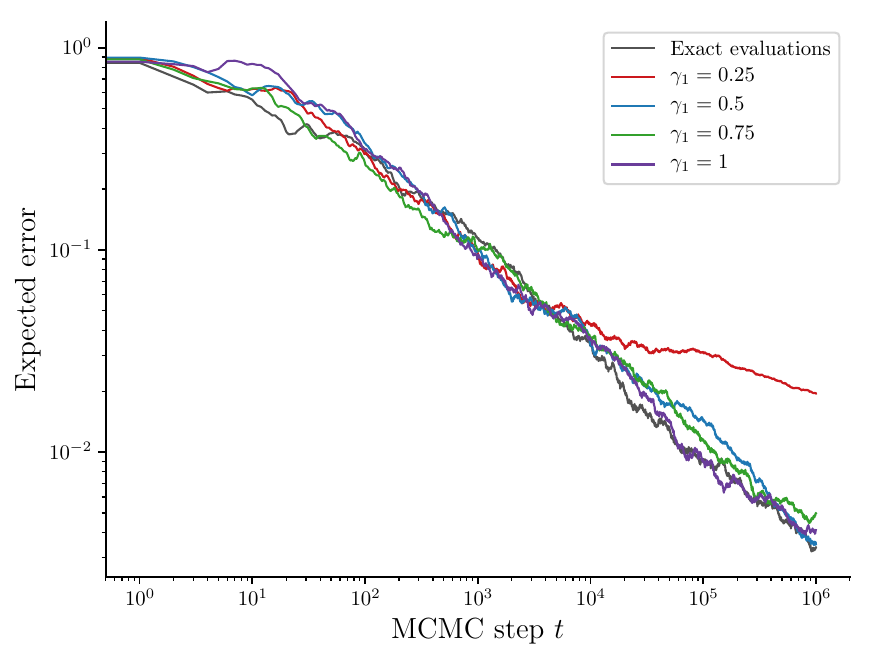} \\[\abovecaptionskip]
    \small (a)
  \end{tabular}

  \vspace{\floatsep}

  \begin{tabular}{@{}c@{}}
    \includegraphics[width=0.475\textwidth]{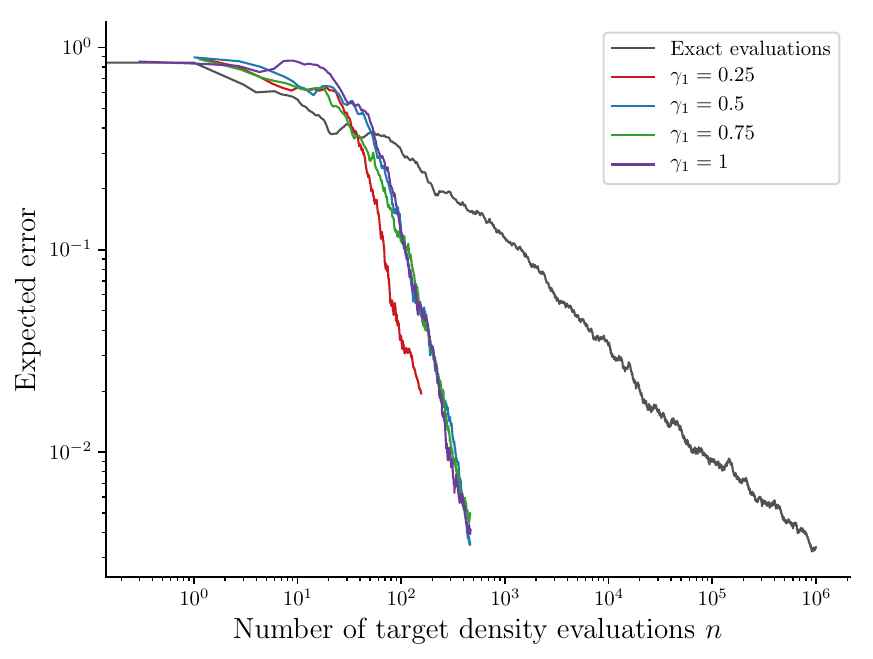} \\[\abovecaptionskip]
    \small (b)
  \end{tabular}

  \caption{The expected errors in variance ($e_t$ and $\breve{e}_t$) averaged over $50$ MCMC chains as a function of (a) MCMC steps $t$ and (b) the number of target density evaluations $n$.  The LA-MCMC construction ensures that, if $\gamma_1  \geq 0.5$, the expected error decays at essentially the same $1/\sqrt{t}$ rate as in the exact evaluation case (Theorem~\ref{thm:convergence-rate}). Since we do not need to evaluate the target density at every MCMC step, and in fact evaluate the target much less frequently over time (see Figure~\ref{fig:1d-example-expected-refinements}), the error decays much more quickly as a function of the number of target density evaluations.}
  \label{fig:1d-example-gamma}
\end{figure}

The bias-variance trade-off renders LA-MCMC insensitive to the error decay rate $\gamma_1$, {as long as} $\gamma_1 \geq 0.5$. This is borne out in Figure \ref{fig:1d-example-gamma}(a), which shows that the error in the variance estimate decays at the same ($1/\sqrt{t}$) rate  as in the exact evaluation case for all values of $\gamma_1$ except $\gamma_1 = 0.25$. 
%
%
Recall that we impose a piecewise constant error threshold $\gamma_0 \ell^{-\gamma_1}$, fixed for each level $\ell(t)$. If $\gamma_1 < 0.5$, then the level lengths $\tau_l$ decrease as $t \rightarrow \infty$; see Section~\ref{sec:bias-variance-trade-off}. In our practical implementation, we can increase $\ell$ at most once per MCMC step and, therefore, when the level length is less than one step, the error threshold cannot decay quickly enough. In this case, the surrogate bias dominates the error and, as we see in Figure \ref{fig:1d-example-gamma}(a), the error decays more slowly as a function of MCMC steps. 

Figure \ref{fig:1d-example-gamma}(b) shows that, since we do not evaluate the target density every MCMC step, the convergence of LA-MCMC as a function of the number of \emph{density evaluations} $n$ is in general much \emph{faster} than in the exact evaluation case.

\begin{figure}
  \centering
  \begin{tabular}{@{}c@{}}
    \includegraphics[width=0.475\textwidth]{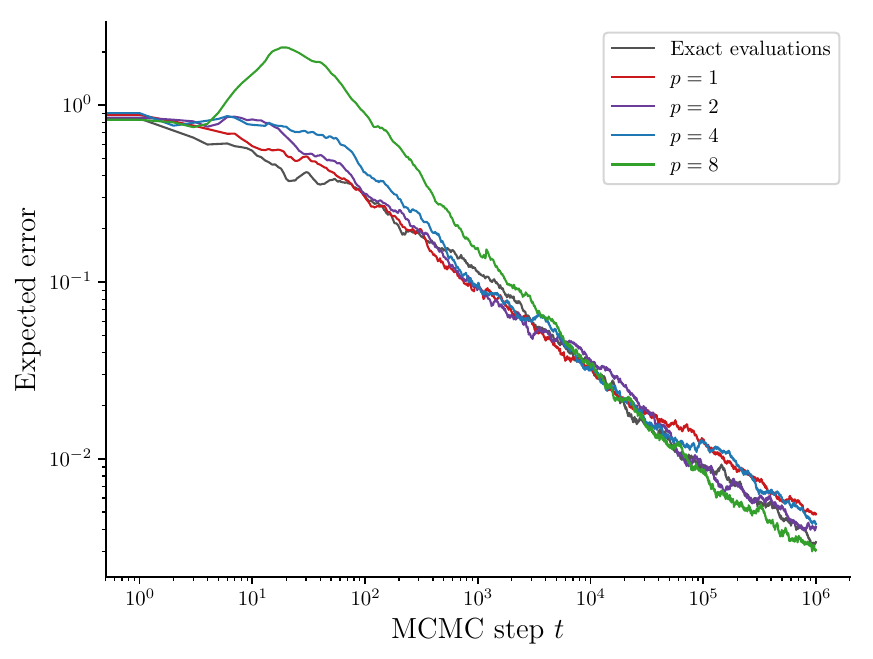} \\[\abovecaptionskip]
    \small (a)
  \end{tabular}

  \vspace{\floatsep}

  \begin{tabular}{@{}c@{}}
    \includegraphics[width=0.475\textwidth]{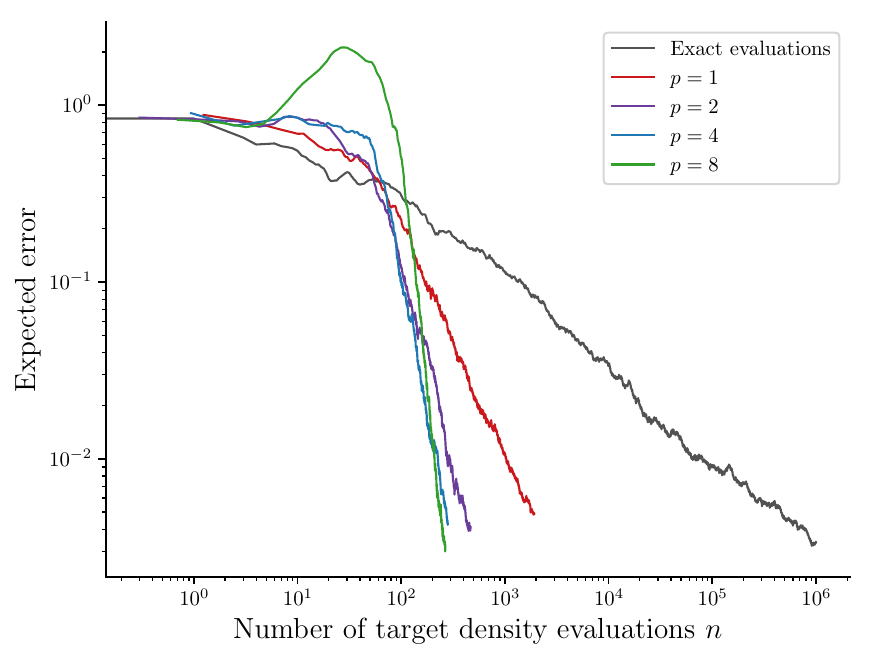} \\[\abovecaptionskip]
    \small (b)
  \end{tabular}

  \caption{The expected errors $e_t$ and $\breve{e}_t$ averaged over $50$ MCMC chains as a function of (a) MCMC steps $t$, and (b) the number of target density evaluations $n$. Controlling the bias-variance trade-off ensures that error decays at essentially the same $1/\sqrt{t}$ rate as in the exact evaluation case. As we increase the order of the local polynomial approximation, however, LA-MCMC requires fewer target density evaluations $n$ to achieve the same error.}
  \label{fig:1d-example-order}
\end{figure}

Approximating the target density with polynomials of higher degree $p$ increases the efficiency of LA-MCMC; we explore this in Figure \ref{fig:1d-example-order}. Figure \ref{fig:1d-example-order}(a) shows that controlling the bias-variance trade-off ensures the error decay rate---as a function of the number of MCMC steps $t$---is the same regardless of $p$. Since the local error indicator is $\Delta(x)^{p+1}$, larger values of $p$ achieve the same error threshold with larger radius $\Delta(x) < 1$. Using higher-degree polynomials therefore requires \emph{fewer} target density evaluations, as shown in Figure \ref{fig:1d-example-order}(b). Yet higher-degree polynomials also require more evaluated points inside the local ball $\mathcal{B}_k(x)$. In this one-dimensional example, the local polynomial requires $p+1$ points to interpolate, and here we choose $k=2(p+1)$ nearest neighbors to solve the regression problem \eqref{eq:local-polynomial-estimate}. We thus see diminishing returns as $p$ increases: higher order polynomials achieve the same accuracy with larger $\Delta(x)$ but require more target density evaluations within each ball $\mathcal{B}_k(x)$. 

\begin{figure}[h!]
\centering
    \includegraphics[width=0.475\textwidth]{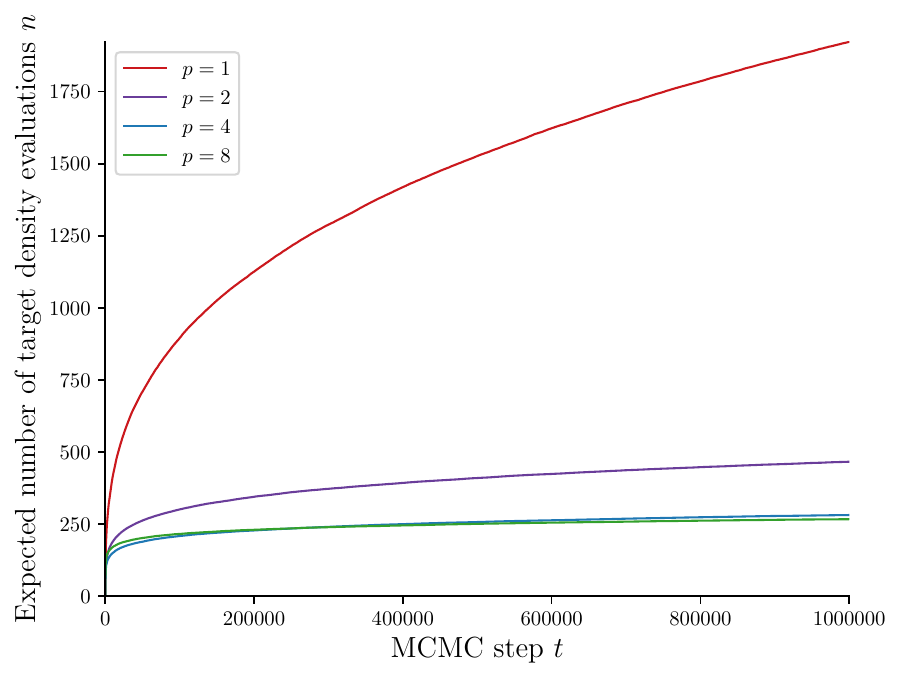}
    \caption{The expected number of refinements $n(t)$, computed from $50$ independent LA-MCMC chains, given different local polynomial degrees $p$.
    The refinement rate decreases as $t \rightarrow \infty$, making MCMC more efficient as $t$ increases.}
    \label{fig:1d-example-expected-refinements}
\end{figure}

As the number of MCMC steps $t \rightarrow \infty$, the rate at which LA-MCMC requires new target density evaluations---i.e., the refinement rate---slows significantly. Figure \ref{fig:1d-example-expected-refinements} illustrates this pattern: the number of target density evaluations increases much more slowly than $t$. While the bias-variance trade-off ensures that the error decay rate remains $1/\sqrt{t}$, viewed in terms of the number of target density evaluations $n$, the picture is different. If target density evaluations are the dominant computational expense---as is typical in many applications---then LA-MCMC generates samples more efficiently as $t \rightarrow \infty$.

\subsection{Controlling tail behavior} \label{sec:example-banana}

\begin{figure}
\centering
  \includegraphics[width=0.475\textwidth]{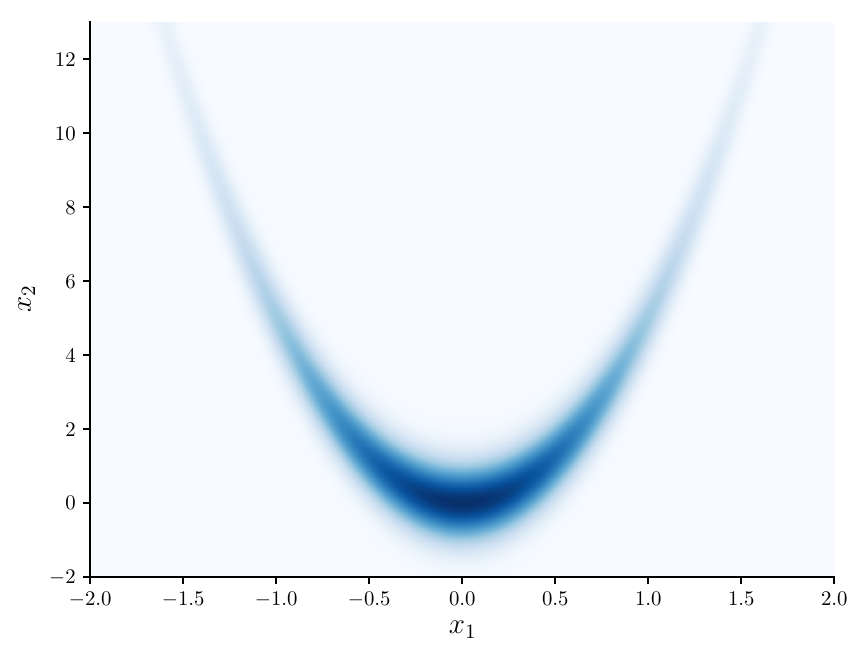} \\[\abovecaptionskip]
  \caption{The two dimensional density defined in \eqref{eq:banana-density}.}
  \label{fig:banana-density-exact}
\end{figure}

\begin{figure}
\centering
  \includegraphics[width=0.475\textwidth]{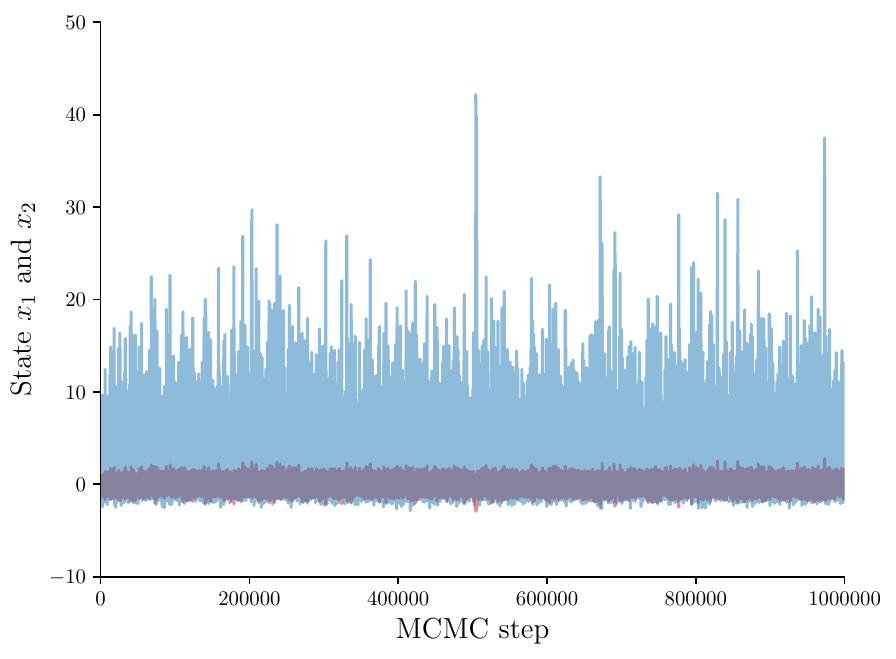} \\[\abovecaptionskip]
  \caption{Trace plots of an MCMC chain targeting \eqref{eq:banana-density}, using exact evaluations.}
  \label{fig:banana-density-exact-mixing}
\end{figure}

The algorithmic parameter $\eta$ (see \eqref{EqQvDef}) controls how quickly LA-MCMC explores the tails of the target distribution. In the previous example, the tails  of the log-target density decayed quadratically---like a Gaussian---and therefore we set $\eta=0$. Now we consider the ``banana-shaped'' density
\begin{equation}
    \log{\pi(x)} \propto -x_1^2 - (x_2-5x_1^2)^2 
    \label{eq:banana-density}
\end{equation}
for $x \in \mathbb{R}^2$. Figure \ref{fig:banana-density-exact}  illustrates this target distribution; note the long tails in the $x_2$ direction. The two traces in Figure \ref{fig:banana-density-exact-mixing} show the mixing of an adaptive Metropolis \citep{Haarioetal2001} chain that uses exact density evaluations. We see that the chain does explore the tails in $x_2$, but always returns to high probability regions; this is the expected and desired behavior of an MCMC algorithm for this target.

\begin{figure}
  \centering

  \begin{tabular}{@{}c@{}}
    \includegraphics[width=0.475\textwidth]{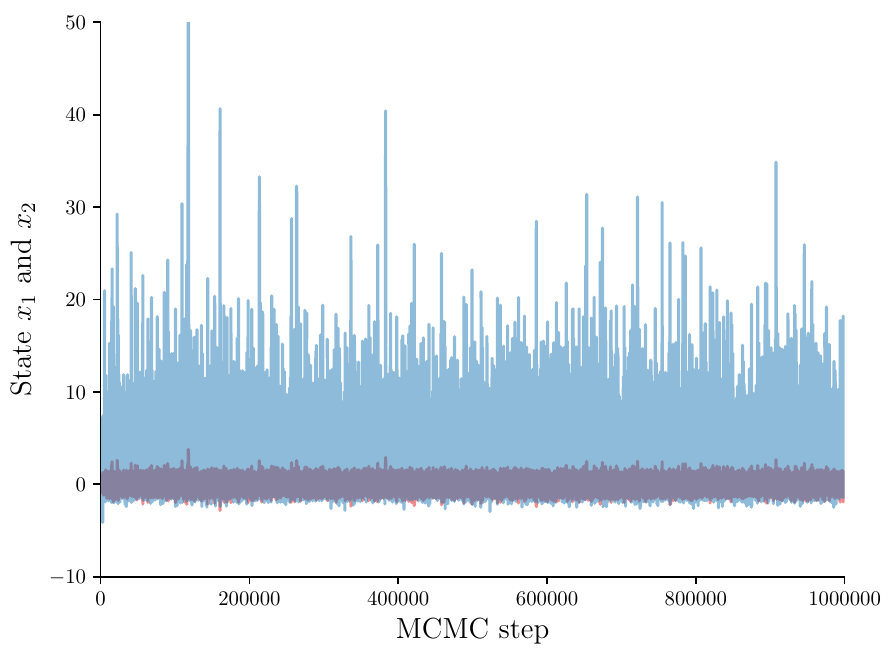} \\[\abovecaptionskip]
    \small (a) $\eta=0.01$
  \end{tabular}
  
  \begin{tabular}{@{}c@{}}
    \includegraphics[width=0.475\textwidth]{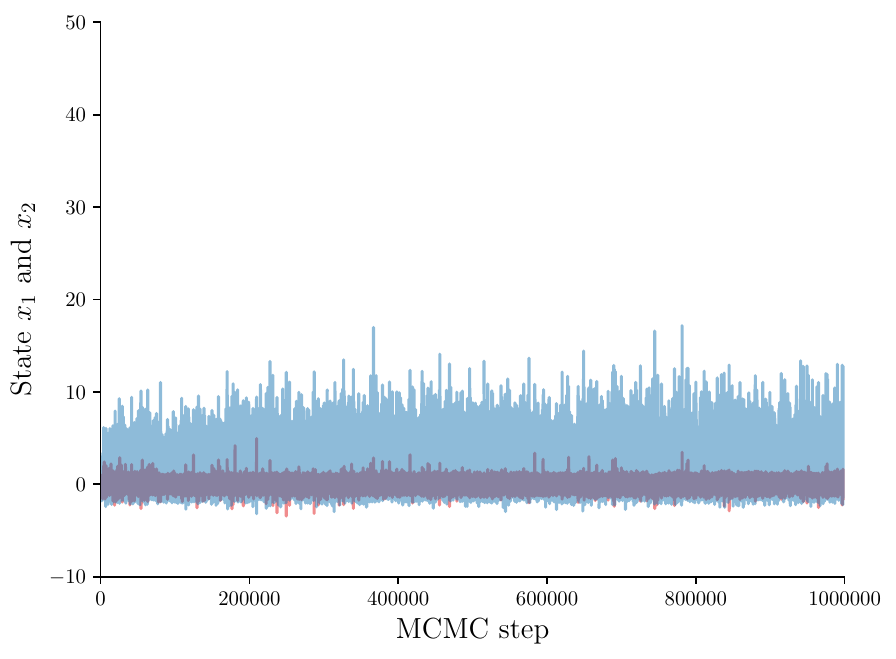} \\[\abovecaptionskip]
    \small (b) $\eta=5$
  \end{tabular}
  
  \begin{tabular}{@{}c@{}}
    \includegraphics[width=0.475\textwidth]{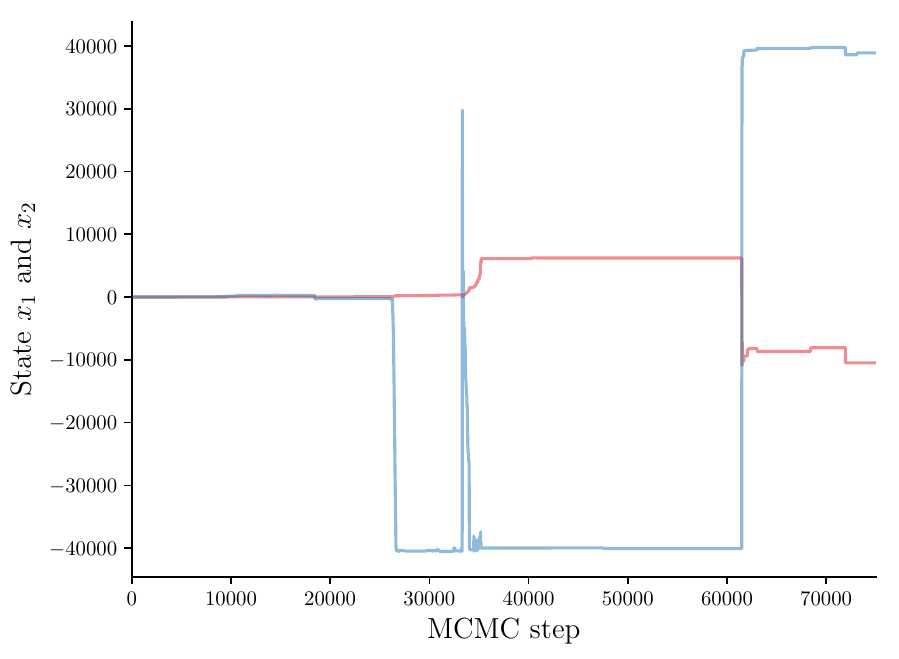} \\[\abovecaptionskip]
    \small (c) $\eta=0$
  \end{tabular}
  
  \caption{Trace plots of LA-MCMC chains targeting \eqref{eq:banana-density} using different values of the tail correction parameter $\eta$. Other algorithmic parameters are $\gamma_0 = 2$, $\gamma_1 = 1$, $\bar{\Lambda} = \infty$, $\tau_0 = 1$, $k = 15$, $p = 2$, and $V(x) = \exp{(0.25 \|x\|^{0.75})}$.}
  \label{fig:banana-density-la-mixing}
\end{figure}

\begin{figure}
  \centering
  
  \begin{tabular}{@{}c@{}}
    \includegraphics[width=0.475\textwidth]{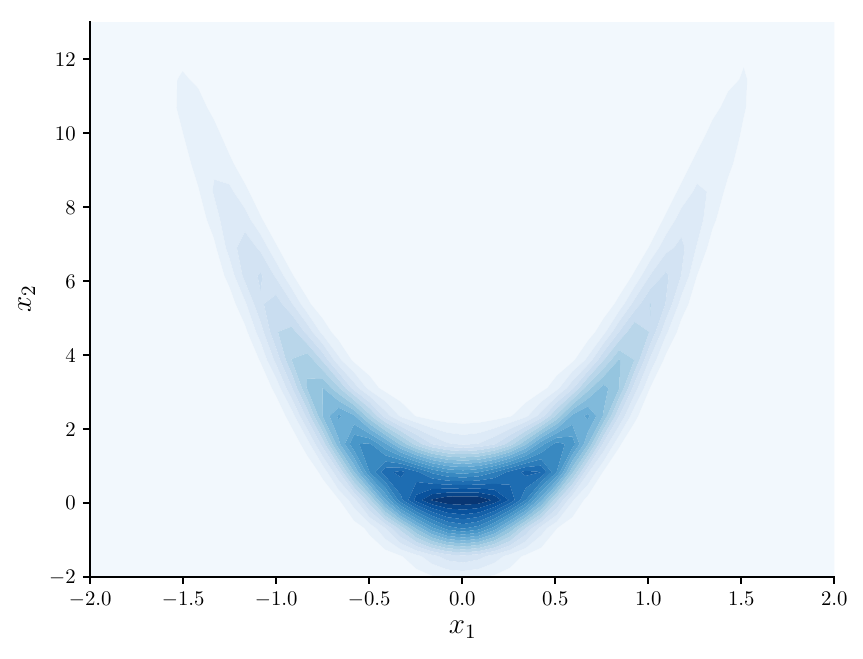} \\[\abovecaptionskip]
    \small (a) $\eta=0.01$
  \end{tabular}

  \begin{tabular}{@{}c@{}}
    \includegraphics[width=0.475\textwidth]{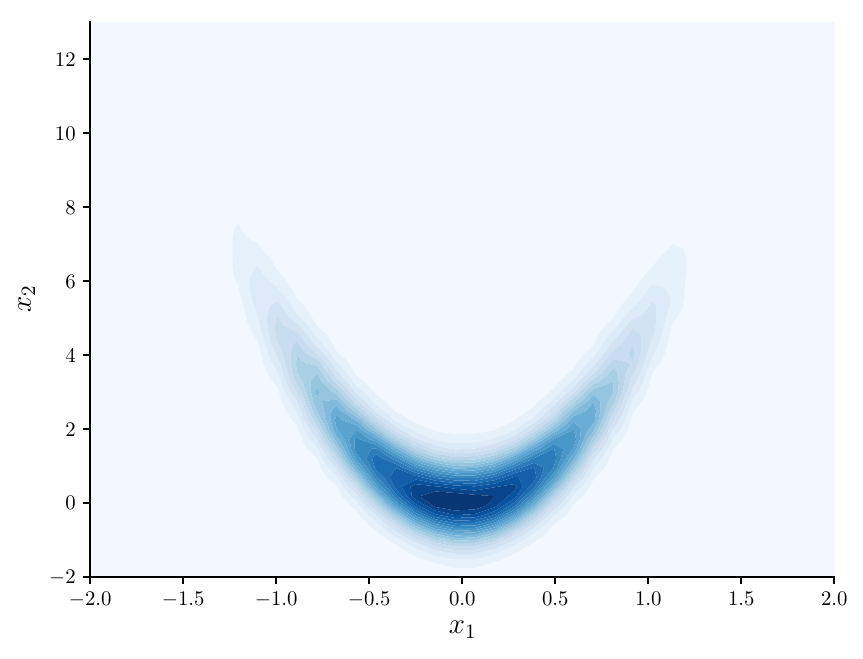} \\[\abovecaptionskip]
    \small (b) $\eta=5$
  \end{tabular}
  
  \caption{Estimates of the target density in \eqref{eq:banana-density} constructed from $10^6$ LA-MCMC samples. We vary $\eta \in \{0.01, 5\}$ and set $\gamma_0 = 2$, $\gamma_1 = 1$, $\bar{\Lambda} = \infty$, $\tau_0 = 1$, $k = 15$, $p = 2$, and $V(x) = \exp{(0.25 \|x\|^{0.75})}$.}
  \label{fig:banana-density-la}
\end{figure}

We can control how quickly LA-MCMC explores the tails of the distribution by varying $\eta$. Here, the local polynomial approximation of $\mathcal{L} = \log \pi$ might not (at any finite time) correctly capture the tail behavior; the surrogate model is therefore, in general, not an unnormalized probability density. 
Using LA-MCMC with no corrections ($\eta=0$) thus allows the chain to wander into the tail without returning to the high-probability region, as shown in Figure \ref{fig:banana-density-la-mixing}(c). Increasing the tail correction parameter $\eta$ biases the acceptance probability so that proposed points close to the centroid are more likely to be accepted and those that are farther are more likely to be rejected. As the number of MCMC steps $t \rightarrow \infty$, this biasing diminishes. The traces in Figure \ref{fig:banana-density-la-mixing}(a) show that setting $\eta > 0$ (here $\eta = 0.01$)  prevents the chain from wandering too far into the tails, and Figure \ref{fig:banana-density-la}(a) shows that the resulting samples correctly characterize the target distribution. 

If $\eta$ is too large, however, then the correction will reduce the efficiency with which the chain explores the distribution's tail. The trace plot in Figure \ref{fig:banana-density-la-mixing}(b) shows that for $\eta=5$, the chain appears to be mixing well. When we compare this chain to Figure~\ref{fig:banana-density-exact-mixing} or Figure~\ref{fig:banana-density-la-mixing}(a), though, we see that the chain is not spending any time in the tail of the distribution. Indeed, the density estimate in Figure \ref{fig:banana-density-la}(b) shows that the tails of the distribution are missing. Asymptotically, the tail correction decays and the algorithm \emph{will} correctly characterize the target distribution for any $\eta > 0$. But too large an $\eta$ can have an impact at finite time. In general, we choose the smallest $\eta$ that prevents the chain from wandering into the distribution's tail.

\subsection{Inferring aquifer transmissivity} \label{sec:tracer}


Now we demonstrate the usefulness of LA-MCMC in \edits{a more computationally demanding example: inferring the spatially heterogeneous transmissivity of an unconfined aquifer. Although this example is physically motivated, our main goal is to highlight important aspects of the LA-MCMC algorithm and its performance. We choose this example because it is related to examples used in previous work \citep{Conradetal2018} and because similar models are used in the groundwater literature---see, for example, \citet{matott2012screening,janettietal2010,al2018,jardani2012,willmann2007,pooletal2015,govTransmissivity200340}. Though our model is idealized, it is not unreasonably different from many models that are used in practice. Moreover, explaining state-of-the-art groundwater models is well beyond the scope of this paper. We refer to existing work (e.g., \citet{janettietal2010}) for a detailed discussion of inferring transmissivity fields in hydrological applications, and here we focus on the computational demonstration of LA-MCMC.
}
In this problem, the likelihood contains a set of coupled partial differential equations that model transport of a nonreactive tracer through the aquifer; the tracer concentration is then observed, with noise, at selected locations in the domain. Each likelihood evaluation is thus computationally intensive. 

\begin{figure}[h!]
\centering
    \includegraphics[width=0.475\textwidth]{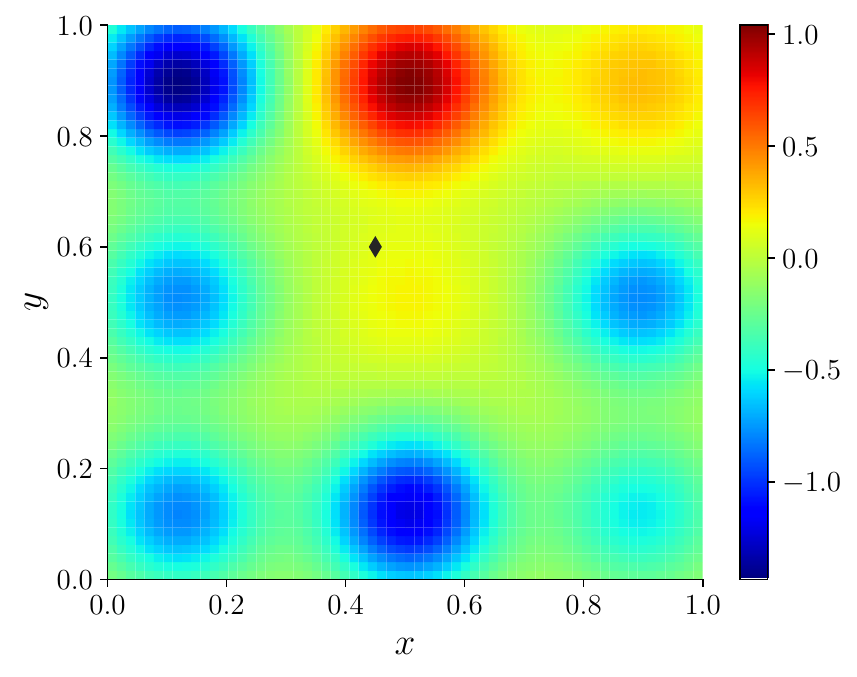}
    \caption{The ``true'' log-transmissivity for an unconfined aquifer model. The coefficients in \eqref{eq:log-transmissivity} are sampled from a standard Gaussian distribution. The diamond denotes a well location.}
    \label{fig:tracer-transport-domain}
\end{figure}

\begin{figure}[h!]
\centering
    \includegraphics[width=0.475\textwidth]{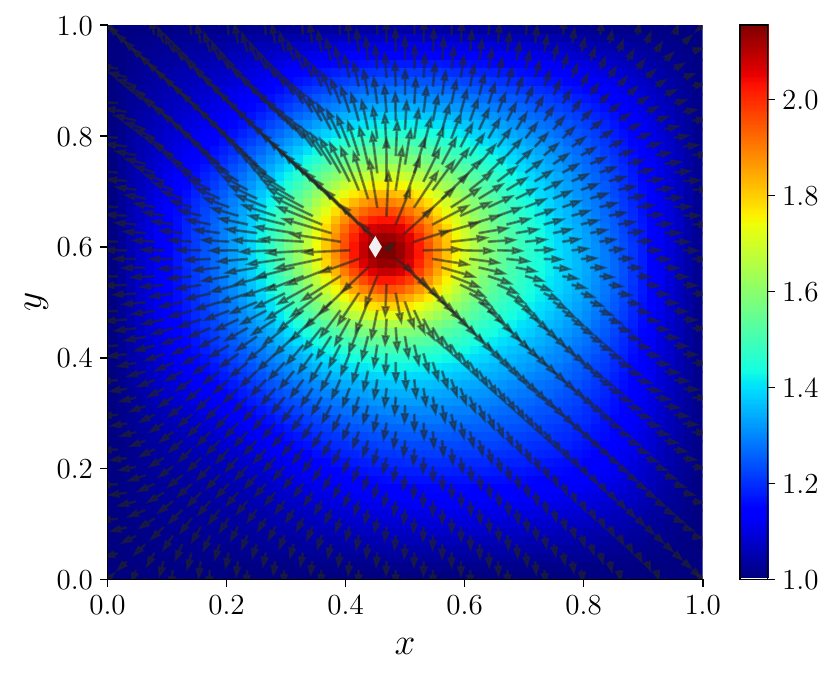}
    \caption{The steady-state hydraulic head and velocity field computed by solving \eqref{eq:hydraulic-head} and \eqref{eq:tracer-velocity} given the ``true'' parameter values are sampled from a standard Gaussian distribution. The white diamond represents the location of a well in the aquifer.}
    \label{fig:tracer-transport-hydraulic-head}
\end{figure}

\begin{figure}[h!]
\centering
    \includegraphics[width=0.475\textwidth]{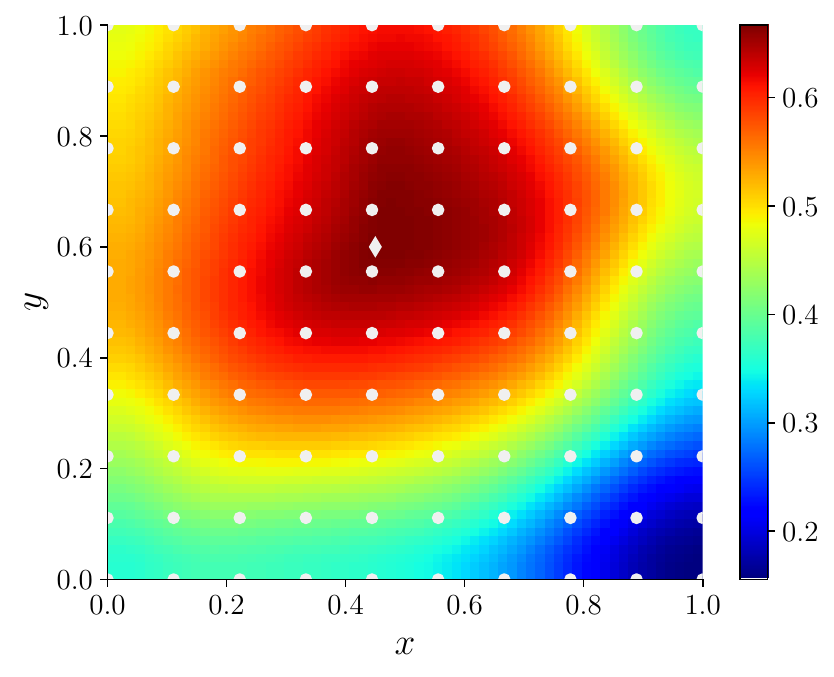}
    \caption{Tracer concentration computed at time $t=1$ by solving \eqref{eq:tracer-concentration} given the ``true'' parameter values in Figure~\ref{fig:tracer-transport-domain} and the steady state velocity field shown in Figure \ref{fig:tracer-transport-hydraulic-head}. The white diamond represents the location of a well in the aquifer, and the circles are sensor locations used in the Bayesian inference problem.}
    \label{fig:tracer-transport-concentration}
\end{figure}

More specifically, the tracer is advected and diffused through the unconfined aquifer (a groundwater resource whose top boundary is not capped by an impermeable layer of rock/soil) by a steady \edits{state} velocity field. 
\edits{
We model the aquifer's log-transmissivity, which determines the permeability of the soil, as a random field on the unit square $\mathcal{D} = [0,1]^2$, parameterized as
\begin{equation}
    \log{ \kappa(z) } = \sum_{i=1}^{d} \kappa_i \sqrt{\lambda_i} e_i(z),
    \label{eq:log-transmissivity}
\end{equation}
where $\kappa_i$ are scalar coefficients, $z \in \mathcal{D}$, and $\{ (\lambda_i, e_i(z))\}_{i=1}^d$ are the $d=9$ leading eigenvalues/eigenfunctions of the integral operator associated with the squared exponential kernel $k(z, z^{\prime}) = \exp{(-\|z-z^{\prime}\|^2/ 2 L^2 )}$, with $L=0.1$. In other words, we have 
\begin{equation}
     \int_{\mathcal{D}} k(z, z^{\prime}) e_i(z^{\prime}) \, dz^\prime = \lambda_i e_i(z) \, .
\end{equation}
Figure~\ref{fig:tracer-transport-domain} shows the ``true'' log-transmissivity, where the true $\kappa_i$ are drawn from a standard normal distribution.}
An ``injection'' well is located at \edits{$(x_w, y_w) = (0.45, 0.6)$}. Given a value of the field $\kappa(z)$, we first compute the steady state hydraulic head $h$ by solving
\begin{subequations}
\begin{equation}
    - \nabla \cdot (\kappa h \nabla h) = f_h
\end{equation}
on the domain $\mathcal{D} \ni z \equiv (x,y)$, with Dirichlet boundary conditions \edits{$h(x,y)=1$ for $x=0$, $x=1$, $y=0$, or $y=1$.} We set the source term above to:
\edits{
\begin{equation*}
    f_h(x,y) = 250 \exp{(-((x-x_w)^2+(y-y_w)^2)/0.005)},
\end{equation*}
}
\label{eq:hydraulic-head}
\end{subequations}
which models hydraulic forcing due to well pumping. The velocity is then
\begin{equation}
    u = -\kappa h \nabla h.
    \label{eq:tracer-velocity}
\end{equation}
The hydraulic head and velocity resulting from the ``true'' parameter values (Figure~\ref{fig:tracer-transport-domain}) are illustrated in Figure~\ref{fig:tracer-transport-hydraulic-head}. Given the steady-state velocity field $u$ and an initial tracer concentration $c(x, y, 0) = 0$, advection and diffusion of the tracer throughout the domain is modeled by a time-dependent concentration field $c(x,y,t)$ that obeys the following transport equation:
\begin{equation}
    \frac{\partial c}{\partial t} + \nabla \cdot ((d_m \mathbf{I} + d_{\ell} u u^T) \nabla c) - u^T \nabla c = -f_t
    \label{eq:tracer-concentration}
\end{equation}
with \edits{$d_m=0.05$, $d_{\ell}=0.001$, and} 
\begin{equation}
    f_t(x,y) = \exp{(-((x-x_w)^2+(y-y_w)^2)/0.005)}.
\end{equation}
The tracer forcing $f_t(x,y)$ models tracer leakage into the domain. The concentration field at time $t = 1$ is shown in \moreedits{Figure \ref{fig:tracer-transport-concentration}}. 

Our goal is to infer the log-transmissivity parameters $(\kappa_i)_{i=1}^d$ given observations of the tracer concentration. We endow each $\kappa_i$  with an independent standard Gaussian prior distribution, $\kappa_i \sim N(0, 1)$. 
\edits{We could also infer other model parameters (e.g., $d_m$, $d_l$, or the forcing functions $f_h$ and $f_t$) by including them in the set of inferred parameters. For the sake of this example, however, we assume that these parameters are known and therefore fixed.}
Data are collected on a $10 \times 10$ array of points evenly spaced in $x \in [0, 1]$ and $y \in [0, 1]$. These locations are marked in Figure~\ref{fig:tracer-transport-concentration}. We make observations at $10$ evenly spaced times $t \in [0, 1]$. 
\edits{This is a relatively large amount of data---typical observation arrays only partially observe the aquifer, although there are often still hundreds of observations (see, e.g., \citet{al2018,janettietal2010}). The large data set here is relatively informative, such that the posterior distribution is quite concentrated relative to the prior and the inference problem is thus more computationally challenging.}
The observations are modeled as
\begin{equation}
    y = f(\kappa) + \varepsilon,
\end{equation}
where $f(\kappa)$ is the \textit{forward model} induced by the partial differential equations above, mapping the log-transmissivity parameters $\kappa \equiv (\kappa_i)_{i=1}^d$ to a prediction of the tracer concentration at the chosen sensor locations/times. The sensor noise is $\varepsilon \sim N(0, 10^{-4} {I})$, yielding the conditional distribution 
\begin{equation}
y \vert \kappa \sim  N(f(\kappa), 10^{-4} I).
\end{equation}
The posterior density then follows from Bayes' rule:
\begin{eqnarray}
    \pi(\kappa \vert y) &\propto& \pi(y \vert \kappa) \pi(\kappa) \, .
    \label{eq:tracer-posterior}
\end{eqnarray}
To avoid a so-called ``inverse crime,'' \edits{where data are generated from the same numerical model used to perform inference \citep{kaipiosomersalo2006},} we generate the data $y$ by solving the forward model with a well-refined $65 \times 65$ numerical discretization using the ``true'' parameters, and then perform the inference using a $25 \times 25$ numerical discretization of the forward model.

Evaluating the forward model at every MCMC step is computationally prohibitive. The run time of the model is \edits{$\mathcal{O}(1)$ seconds on an Intel Core i7-7700 CPU at 3.60GHz, with some variability depending on the value of log-transmissivity parameters $\kappa$.} Generating \edits{$10^6$ samples thus takes 5--10 days} 
of computation time. LA-MCMC is essential to making this Bayesian inference problem computationally feasible. 
\edits{We note that high performance and parallel computing resources could certainly reduce the computational cost of each model evaluation. Parallel computing can even enable the shared construction of surrogate models using concurrent chains, as described in our previous work \citep{Conradetal2018}. Since our goal here is to demonstrate the impact of new local approximation strategies, however, we focus on a serial implementation. A useful hardware- and implementation-independent measure of computational cost is the number of refinements (i.e., expensive model/target density evaluations) in a given run, which we discuss below.
} 


Recall that the local polynomial surrogate \eqref{eq:local-polynomial-estimate} requires computing $q = \text{dim}(\mathcal{P})$ coefficients, and in the total-degree setting of our theory, we have $q = {{d+p}\choose{p}}$. This quantity grows rapidly with the dimension $d$ of the parameters for $p>1$.
Retaining a total-degree construction, we could set $p=0$ and thus include only a constant term in $\mathcal{P}$. This results in a constant surrogate model within each ball; now the number of terms in the local polynomial approximation is always one, independent of the parameter dimension. Similarly, setting $p=1$ lets $q$ grow only linearly with $d$. Yet in the one-dimensional example of Section~\ref{sec:1d-example}, we showed that including higher-degree terms in the approximation can reduce the number of expensive target density evaluations required to achieve a given accuracy (see Figure \ref{fig:1d-example-order}).

We implement these total-degree local polynomial approximations but also compare them empirically with other choices of polynomial approximation space---in particular, truncations that retain higher-order terms more selectively. A common practice in high-dimensional  approximation is to employ \emph{sparse truncations} of the relevant index set \citep{BlatmanSudret2011}. Let $\psi_s(\kappa_i)$ be a univariate polynomial of degree $s$ in $\kappa_i$. (Typically, we choose an orthogonal polynomial family, but \moreedits{this choice} is immaterial.) Each $d$-variate polynomial basis function $\phi_{\bm{\alpha}}(\kappa) \coloneqq \prod_{i=1}^{d} \psi_{\alpha_i}(\kappa_i)$ is defined by a multi-index $\bm{\alpha} \in \mathbb{N}_0^d$, where the integers $\alpha_i \geq 0$ are components of $\bm{\alpha}$. Now define the $\ell^\nu$ norm for $\nu > 0$ (a quasi-norm for $0 < \nu < 1$): 
\begin{equation}
    \|\bm{\alpha}\|_{\nu} = \left(\sum_{i=1}^{d} \alpha_i^{\nu}\right)^{1/\nu}.
\end{equation}
Given a maximum degree $p$, we can build our \emph{local} polynomial approximations in the polynomial space $\mathcal{P}_\nu^{p,d}$ spanned by basis functions with multi-indices $\|\bm{\alpha}\|_{\nu} \leq p$. When $\nu=1$, this recovers a total-degree expansion, which is consistent with the error bounds in Section~\ref{sec:erroranalysis}. However, the dimensionality of $\mathcal{P}_\nu^{p,d}$ decreases as $\nu \rightarrow 0$. We define the special case of $\nu=0$ to be a polynomial space with no cross terms: each multi-index has one nonzero element, and that element is at most $p$. 

\begin{figure}[h!]
\centering
    \includegraphics[width=0.475\textwidth]{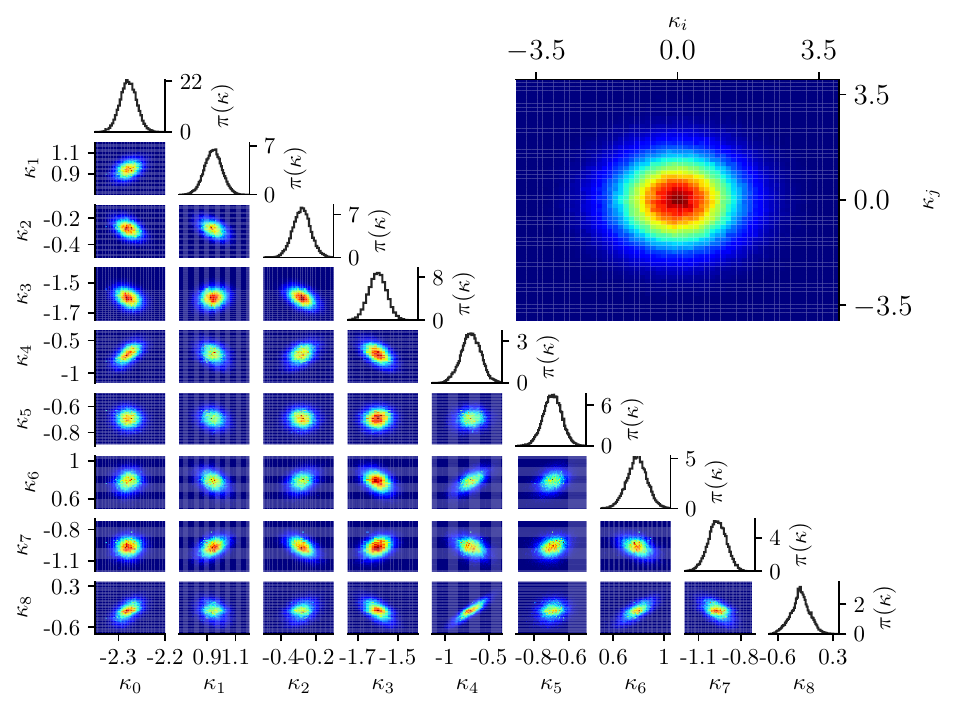} 
    \caption{The diagonal and lower diagonal subplots show the one- and two-dimensional marginal distributions of the posterior distribution \eqref{eq:tracer-posterior} estimated using LA-MCMC with \edits{$\nu=0.75$, $\gamma_0=2$, $\gamma_1=0.5$, $\bar{\Lambda}=\infty$, $\tau_0 = 1$, $\eta = 0$, $k=75$, $p=3$, and $V(x) = \exp{(\|\kappa-\kappa_{\text{MAP}}\|)}$. Here, $\kappa_{\text{MAP}} = \argmax_{\kappa}{\pi(\kappa \vert y)}$ is the posterior mode. The subplot on the upper right shows the two-dimensional prior marginal for any pair of parameters.}}
    \label{fig:tracer-transport-marginals}
\end{figure}

\begin{figure}
  \centering
  
  \begin{tabular}{@{}c@{}}
    \includegraphics[width=0.475\textwidth]{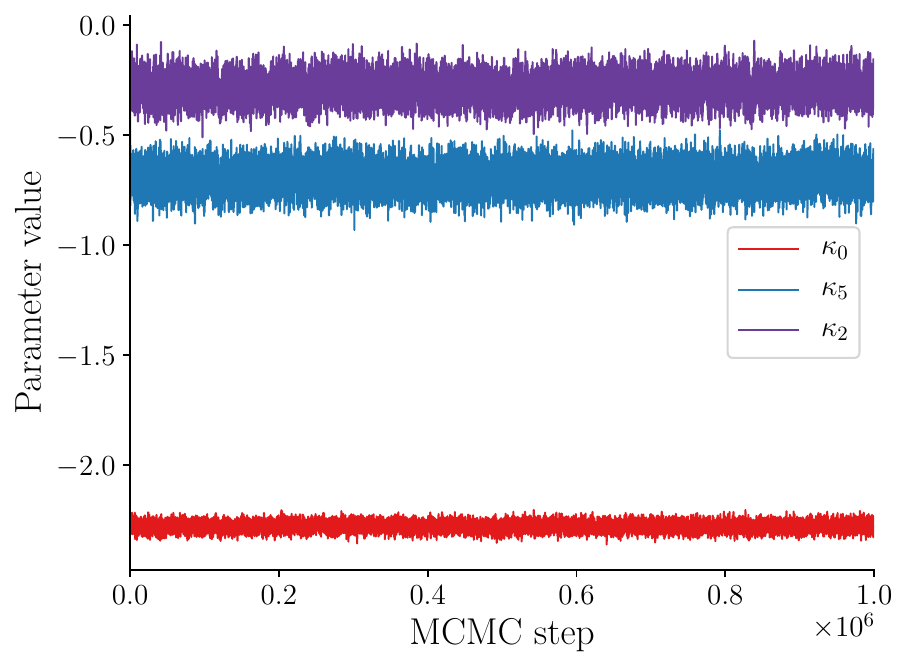} \\[\abovecaptionskip]
    \small (a) Exact MCMC mixing
  \end{tabular}
  
  \begin{tabular}{@{}c@{}}
    \includegraphics[width=0.475\textwidth]{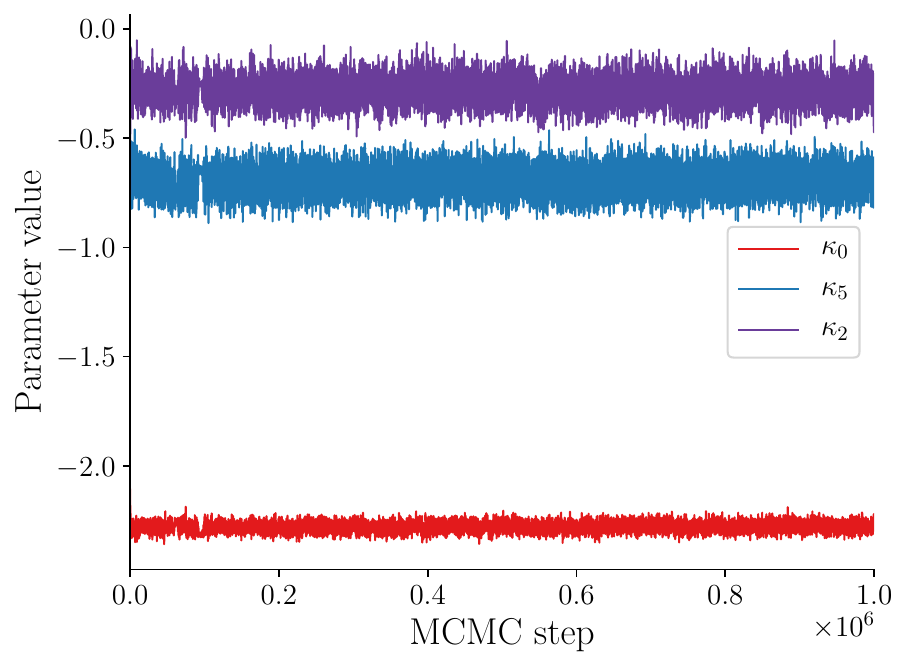} \\[\abovecaptionskip]
    \small (a) LA-MCMC mixing
  \end{tabular}
  
  \caption{PDE/tracer transport problem: trace plot of \edits{three parameters of (a) an exact MCMC chain and (b) an LA-MCMC chain using a locally cubic ($p=3$) surrogate model with $\nu = 0.75$.} Additional parameters are defined on the last row of Table \ref{tab:tracer-mcmc-parameters}.}
  \label{fig:tracer-mixing}
\end{figure}

We now apply the LA-MCMC algorithm, \edits{as well as an exact MCMC algorithm for comparison,} to the posterior distribution of the tracer transport problem. \edits{All MCMC chains use the same (fixed) random-walk proposal $q_t = \mathcal{N}(\kappa_t, 0.005 I)$, where $\kappa_t$ is the current state of the chain.} We tested different configurations of LA-MCMC, parameters of which are given in Table~\ref{tab:tracer-mcmc-parameters}. The Lyapunov function $V$ used in the tail correction is defined using the posterior mode, $\bar{\kappa} = \argmax_{\kappa}{\pi(\kappa \vert y)}$, \edits{which we obtain by maximizing the log-posterior density using optimization algorithms in \texttt{nlopt} \citep{johnson2014nlopt}}. The rows in Table~\ref{tab:tracer-mcmc-parameters} with $\nu =1$ correspond to using total-degree linear and quadratic local polynomial approximations (consistent with our theory). Setting $\nu < 1$ defines a sparse approximation. The required number of nearest neighbors $k$ is chosen to be slightly more than the number of points $q = \text{dim}(\mathcal{P}_\nu^{p,d}) $ required to interpolate. 
Reducing $\nu$ reduces $q$ and $k$. Chains produced by all these algorithmic configurations yield essentially identical posterior estimates. A trace plot of selected states from the LA-MCMC chain corresponding to the $p=3$ configuration is shown in  Figure \ref{fig:tracer-mixing}, \edits{and compared to a trace plot of the exact-model chain}. One- and two-dimensional marginals of the posterior distribution are shown in Figure \ref{fig:tracer-transport-marginals}. While this figure is generated with $\nu=0.75$ and $p=3$, the results are essentially identical for the other algorithmic configurations. \edits{An exact chain of $10^6$ steps took about $6$ days to complete, whereas the LA-MCMC chains of the same length with quadratic and cubic surrogate models took roughly $6$ to $10$ hours, and with linear models took just over one day. These timings are somewhat imprecise, however, as other computational tasks were competing for resources on the same workstation (a Intel Core i7-7700 CPU at 3.60GHz) used for our runs.} 

\begin{table}[h!]
    \centering
    \begin{tabular}{c|c|c|c|c}
        degree $p$ & $\nu$ & $k$ & \# to interpolate ($q$) & $\gamma_0$ \\ \hline \hline
        Exact & --- & --- & --- & ---\\
        $1$ & $1$ & $20$ & $10$ & $1.9$ \\
        $2$ & $1$ & $65$ & $55$ & $1$ \\
        $3$ & $0.75$ & $75$ & $64$ & $1$ 
    \end{tabular}
    \caption{Parameter configurations for Algorithm \ref{alg:la-mcmc}, used to generate samples from the posterior distribution \eqref{eq:tracer-posterior} of the PDE/tracer transport problem. In all cases $\gamma_1 = 0.5$, $\eta = 0$, $\log{V(\kappa)} = \|\kappa-\kappa_{\text{MAP}}\|$, and $\bar{\Lambda} = \infty$. \edits{Here, $\kappa_{\text{MAP}} = \argmax_{\kappa}{\pi(\kappa \vert y)}$ is the posterior mode, which we obtain by maximizing the log-posterior density using \texttt{nlopt} \citep{johnson2014nlopt}}.}
    \label{tab:tracer-mcmc-parameters}
\end{table}

\begin{figure}
  \centering
  
  \begin{tabular}{@{}c@{}}
    \includegraphics[width=0.475\textwidth]{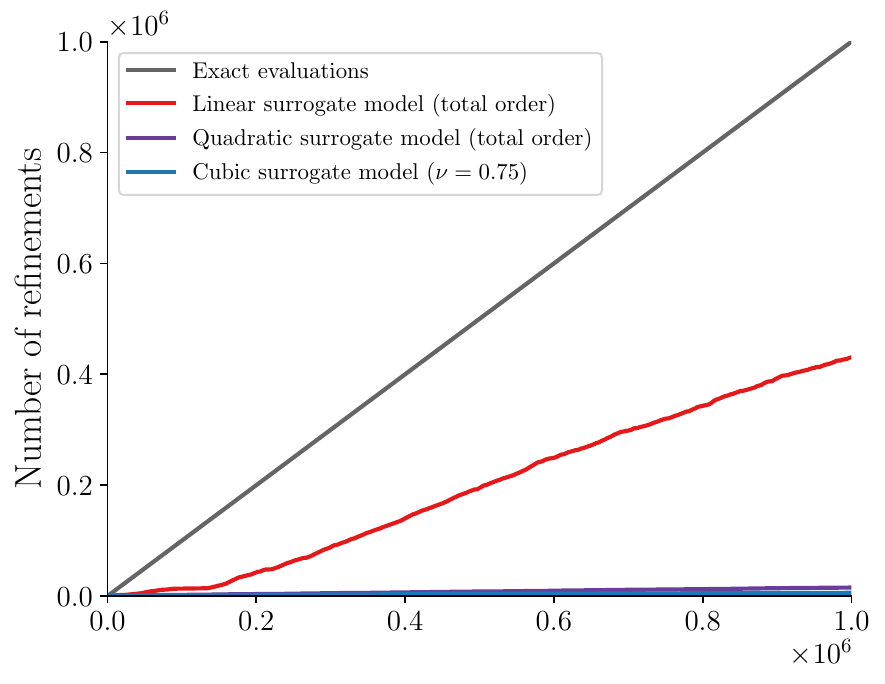} \\[\abovecaptionskip]
    \small (a) Refinements per MCMC step 
  \end{tabular}
  
  \begin{tabular}{@{}c@{}}
    \includegraphics[width=0.475\textwidth]{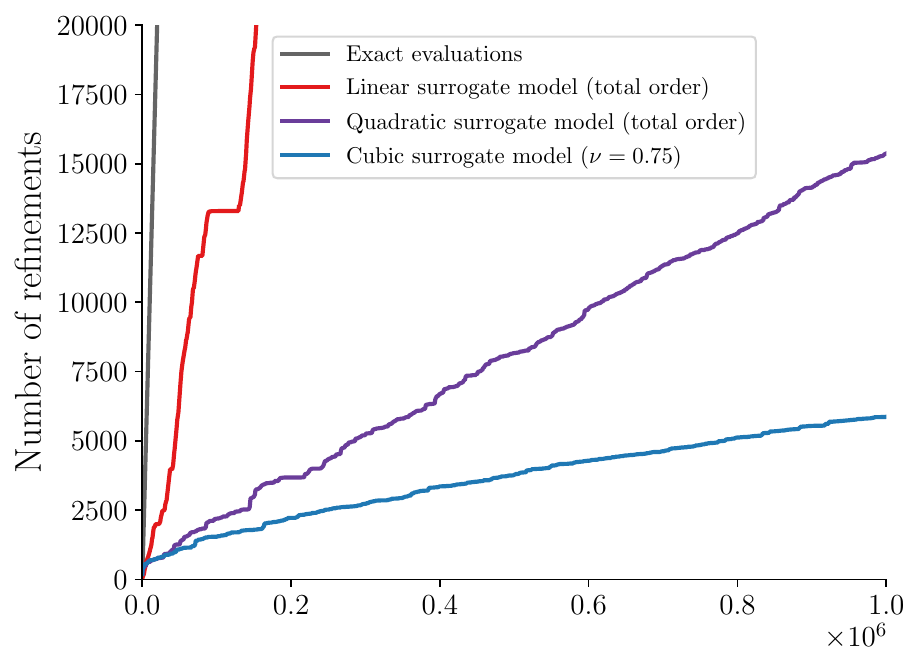} \\[\abovecaptionskip]
    \small (b) Refinements per MCMC step (rescaled)
  \end{tabular}
  
  \caption{PDE/tracer transport problem: number of likelihood evaluations (i.e., refinements) as a function of MCMC steps, for LA-MCMC chains with parameters defined in Table \ref{tab:tracer-mcmc-parameters}. Panel (a) shows the total number of refinements for all chains, and panel (b) rescales the vertical axis to emphasize how many fewer model evaluations LA-MCMC requires than the exact case. For local surrogate models with $p \geq 2$, we see significant reductions (of three to four orders of magnitude) over the number of likelihood evaluations required by exact MCMC.}
  \label{fig:tracer-refinements}
\end{figure}

Though all these chains are successful in characterizing the posterior distribution, the required number of (expensive) likelihood evaluations---and hence the overall computational cost---depend on a non-trivial relationship between the accuracy of the local approximations and the number of nearest neighbors required to define them. 
\edits{As noted earlier, though, the number of such likelihood evaluations is a robust and translatable measure of computational cost, independent of implementation and hardware, and also most meaningful in the setting where model evaluations dominate all other costs of the MCMC machinery.} 
Figure~\ref{fig:tracer-refinements} shows the number of likelihood evaluations $n$ as a function of the number of MCMC steps $t$ for each experiment in Table \ref{tab:tracer-mcmc-parameters}. \edits{We see that the quadratic and cubic local approximations (purple and blue lines) require far fewer expensive likelihood evaluations. In all cases, we see a reduction relative to the number of likelihood evaluations that would be required by exact MCMC. However, this improvement is drastically increased, by orders of magnitude, when choosing $p>1$.}

It is important to note that our approach to controlling the bias-variance tradeoff, and thus the error thresholds for triggering refinement via \eqref{eq:local-error-bound} and \eqref{eq:error-threshold}, are derived for expansions of total degree $p$ and thus in principle applicable only for $\nu=1$. We use the same rules here for $\nu<1$, but this approach cannot guarantee that bias and variance decay at the same rate. Indeed, a full analysis of the local sparse approximations may depend on understanding the magnitudes of mixed derivatives of the target function $g$; this can become quite problem-specific, and we defer such an investigation to future work. The empirical results for $\nu<1$ here are intended to be practical and exploratory.

\begin{figure}
  \centering

  
  \begin{tabular}{@{}c@{}}
    \includegraphics[width=0.475\textwidth]{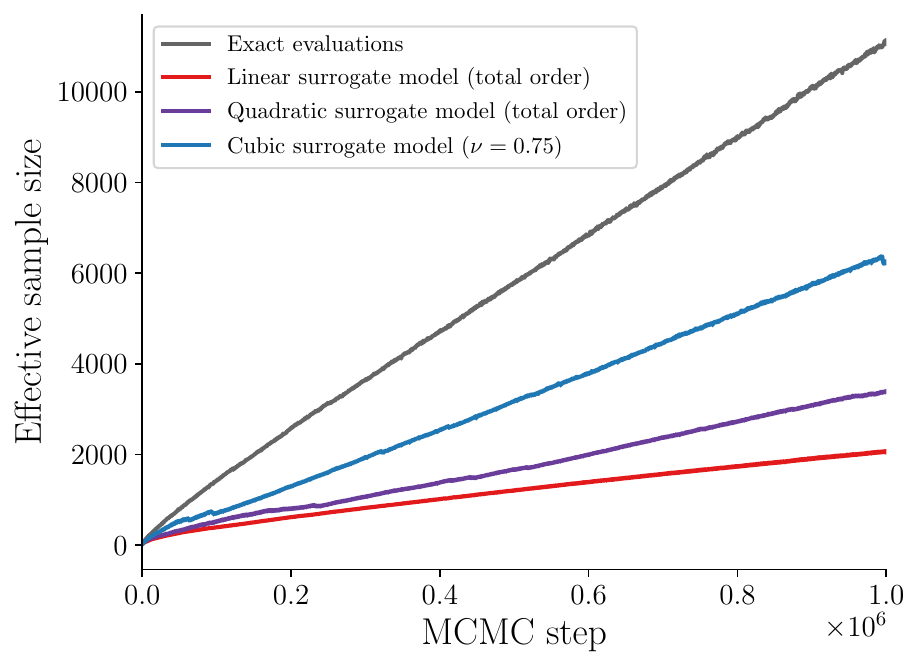} \\[\abovecaptionskip]
    \small (a) ESS per MCMC step 
  \end{tabular}
  
  \begin{tabular}{@{}c@{}}
    \includegraphics[width=0.475\textwidth]{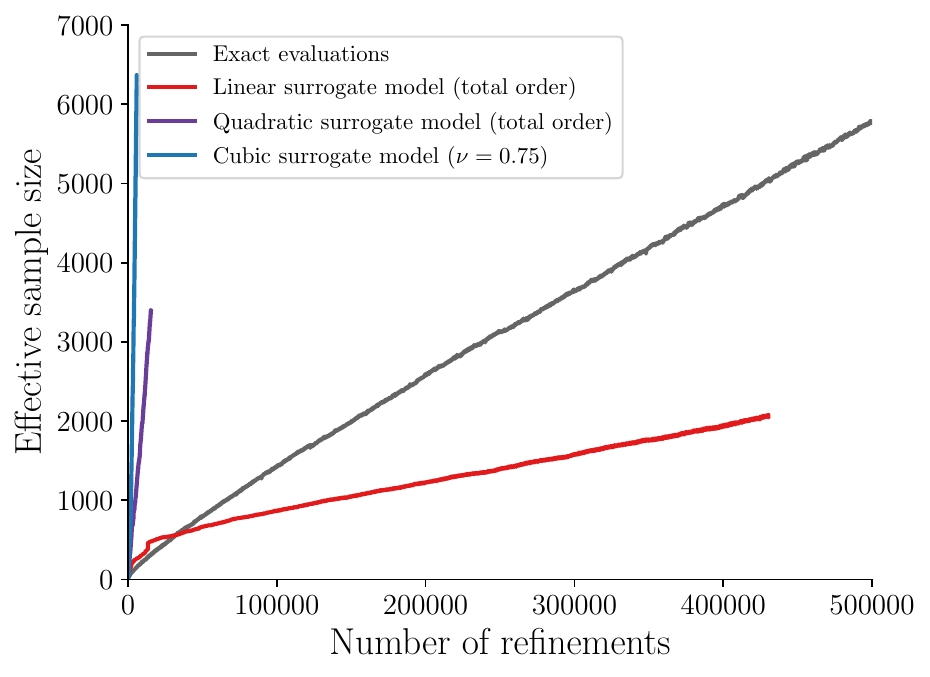} \\[\abovecaptionskip]
    \small (b) ESS per model evaluation
  \end{tabular}

    \begin{tabular}{@{}c@{}}
    \includegraphics[width=0.475\textwidth]{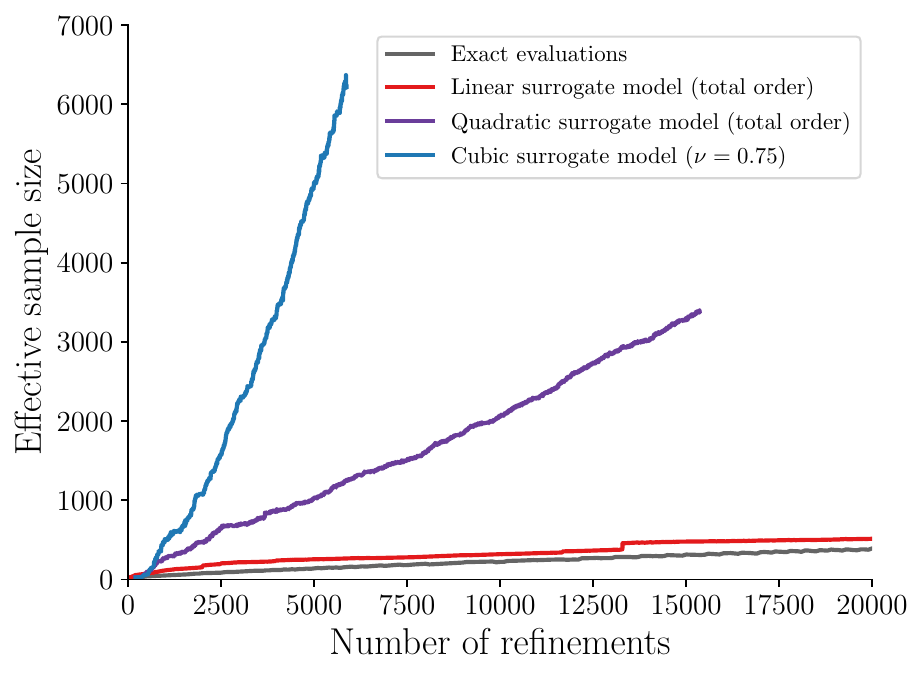} \\[\abovecaptionskip]
    \small (c) ESS per model evaluation (rescaled)
  \end{tabular}
  
  \caption{PDE/tracer transport problem, computational efficiency for LA-MCMC chains with parameters defined in Table \ref{tab:tracer-mcmc-parameters}. (a) effective sample size (ESS) as a function of the number of MCMC steps; (b-c) ESS as a function of the number of likelihood evaluations (refinements). Panel (c) rescales the horizontal axis to more clearly show how quickly LA-MCMC with cubic and quadratic surrogate models generates independent samples as a function of the number of refinements.}
  \label{fig:tracer-autocorrelation-ess}
\end{figure}

The choice of local approximation not only affects the overall computational cost of each chain, but also its mixing. Figure~\ref{fig:tracer-autocorrelation-ess}(a) shows the number of effectively independent samples (i.e., the effective sample size (ESS) \citep{Wolffetal2004}) produced by each configuration, as a function of the number of MCMC steps $t$. \edits{The chain using exact density evaluations generates effectively independent samples fastest \textit{as a function of $t$}. Mixing is slightly faster with cubic local approximations than with local quadratic or linear approximations. Figures \ref{fig:tracer-autocorrelation-ess}(b)--(c), however, show a non-trivial relationship between complexity of the surrogate model and how quickly it generates ESS \textit{as a function of the number of model evaluations $n$}.} Chains employing local linear approximations ($p=1$) \edits{(red line) do not mix as efficiently as the sparse local cubic approximation or the total-degree quadratic approximation (blue and purple lines). Because the number of model evaluations in the $p=1$ case is not so drastically reduced over the exact case (see Figure~\ref{fig:tracer-refinements}(a)), poorer mixing leads to worse performance in the overall metric of ESS per model evaluation. On the other hand, the $p=3$ case achieves the same ESS with 5000 model evaluations that the exact chain achieves with $5\times 10^5$ evaluations, an improvement of two orders of magnitude.

Recall that Figure~\ref{fig:tracer-transport-marginals} shows strong posterior correlations for this problem. We speculate that log-likelihood approximations that include cross terms (i.e., $p>1$) can more easily approximate the true posterior density and, therefore, significantly outperform local approximations that do not include these factors.} 
%
%
In theory, even a locally constant model will yield convergence as $\Delta \to 0$. \edits{In practice, we find that the choice of basis for the local approximation may have a substantial impact. While $p=2$ or $p=3$ are good default choices, a method for adaptively selecting polynomial basis functions as we learn about correlations in the posterior could possibly improve the algorithm. We leave this to future work.
}

\section{Conclusions}
\label{sec:concs}
This paper investigated the design and convergence properties of local approximation MCMC (LA-MCMC) algorithms, which exploit regularity of the target density to build and refine local polynomial surrogate models during sampling. When density evaluations are expensive, these algorithms can reduce the computational cost of MCMC by orders of magnitude.

We introduce a new LA-MCMC algorithm with two key features: (1) a rate-optimal refinement strategy that balances the decay of surrogate-induced bias with the decay of Monte Carlo variance; and (2) a modification to the Metropolis acceptance step of the algorithm, which helps ensure stability of the resulting chain.
Together these features enable broader applicability and stronger theoretical guarantees than previous efforts. Earlier versions of LA-MCMC \citep{Conradetal2016} employed a cross-validation heuristic in conjunction with random refinements of the surrogate model; the latter made possible a guarantee of ergodicity on compact state spaces or under strong tail conditions. In the present effort, we dispense with both cross-validation and random refinement, using instead the more principled bias-variance control strategy mentioned above. Not only is the resulting sampling method asymptotically exact; here we also show that the error in LA-MCMC estimates decays at approximately the expected $1/\sqrt{T}$ rate, where $T$ is the number of MCMC steps. Moreover, this rate holds without the restrictive tail conditions that were required, both in theory and in practice, for stability and convergence of previous versions of the algorithm. 
Using our refinement strategy, we also observe that the rate at which new target density evaluations are demanded decays with $T$. This means that as a function of the number of target density evaluations, convergence is \textit{faster} than that of standard geometrically ergodic MCMC, accelerating as $T$ increases, in accordance with the underlying function approximation machinery.


Our numerical examples demonstrate both the predicted convergence rates and stability of the sampler for target distributions with heavier tails. We also show how LA-MCMC can be employed in a computationally intensive inference problem, where evaluations of the likelihood require solving a coupled set of nonlinear partial differential equations; in this example, motivated by inverse problems in groundwater hydrology, LA-MCMC reduces the number of expensive likelihood evaluations by roughly two orders of magnitude. 

While the present algorithm is well suited to expensive target densities on parameter spaces of moderate dimension (e.g., $d=9$ in the current PDE example), \moreedits{it becomes more challenging to apply in higher dimensions, even $d=15$ in our experience. The fundamental reason is that generic isotropic local polynomial approximation is subject to the curse of dimensionality, with at least ${{d+p}\choose{p}}$ nearest neighbors required to evaluate the surrogate at a given point. Hence a larger number of  target density evaluations are required as $d$ increases, and the benefits of function approximation over sampling begin to diminish. As with any function approximation problem, the key to handling higher dimensions is to exploit some form of structure, e.g., anisotropy or sparsity. For instance,} future work could investigate the use of sparse local polynomial approximations as way of reducing the complexity of the local surrogate in higher dimensions. The preliminary empirical study in this paper shows that sparse local polynomial approximation can be helpful, but further analysis---perhaps building on the analysis of \emph{global} sparse polynomial approximations of likelihood functions in Bayesian inverse problems \citep{schillings2014sparsity} \moreedits{and exploiting anisotropy}---is needed to understand the convergence properties of such approximations within LA-MCMC. \moreedits{Other methods of explicit dimension reduction \citep{zahm2018certified} may also be useful in this setting: the idea would be to find a low-dimensional subspace in which the posterior departs most strongly from the prior and to apply our LA-MCMC machinery to the log-likelihood function only within that subspace. This approach has antecedents in \citet{Cui2016etal}. In the present paper, however, we have focused on the essential approximation and sampling machinery of LA-MCMC and on the analysis of its convergence. Combinations with other techniques, while practically important, are left to future work.}




An open source implementation of the LA-MCMC algorithm introduced in this paper is available as part of the MUQ library (\url{http://muq.mit.edu}).

\appendix

\section{Theoretical results} \label{app:theory}

We include all theoretical results from the paper. Throughout this section, we only deal with the case that the underlying proposal distribution $q_{t}$ does not change with $t$. To deal with typical small adaptations, we believe that the following framework can be combined with, e.g., the approach of \citet{roberts2007coupling}, but this would result in a significantly longer paper and these adaptations are not central to our approach.

\subsection{General bounds on non-Markovian approximate MCMC algorithms}

Proceeding more formally, let $\{\hat{X}_t, \hat{K}_t, \mathcal{F}_t\}_{t \geq 0}$ be a triple satisfying:
\begin{enumerate}
    \item $\{\hat{X}_t\}_{t \geq 0}$ is a sequence of \textit{random variables} on $\mathbb{R}^{d}$;
    \item $\{\hat{K}_t\}_{t \geq 0}$ is a (typically random) sequence of \textit{transition kernels} on $\mathbb{R}^{d}$;
    \item $\{\mathcal{F}_t\}_{t \geq 0}$ is a \textit{filtration}, and $\{\hat{X}_t, \hat{K}_t\}_{t \geq 0}$ is \textit{adapted} to this filtration;
    \item the three agree in the sense that 
    \begin{equation} 
        \mathbb{P}[X_{s+1} \in A \vert \mathcal{F}_{s}] = \hat{K}_s(\hat{X}_s, A)
    \end{equation}
    for all $s \geq 0$ and all measurable $A$. Note that, in particular, both left- and right-hand sides are $\mathcal{F}_s$-measurable random variables in $[0, 1]$.
\end{enumerate}
In practice, $\mathcal{F}_{s}$ is generated by our sequence of approximations to the true log-target. We use the following quantitative assumptions:
\begin{assumption}
(Lyapunov inequality). There exists $V: \mathbb{R}^{d} \to [1, \infty)$ and constants $0 < \alpha \leq 1$ and $0 \leq \beta < \infty$ so that 
\begin{equation}
    (\hat{K}_s V)(\hat{X}_{s}) \leq (1-\alpha) V(\hat{X}_{s}) + \beta
\end{equation} 
and
\begin{equation}
    (KV)(x) \leq (1-\alpha) V(x) + \beta
\end{equation} 
for all $s \geq 0$. The second inequality should hold deterministically; note that this is an $\mathcal{F}_{s}$-measurable event.
\label{assumption:lyapunov-inequality}
\end{assumption}
\begin{assumption}
(Good approximation). Let Assumption \ref{assumption:lyapunov-inequality-simple} or \ref{assumption:lyapunov-inequality}  hold. There exists a monotonically decreasing function $\delta:[0, \infty) \to [0, 0.5)$ so that 
\begin{equation}
    \|K(\hat{X}_{s}, \cdot) - \hat{K}_s(\hat{X}_{s}, \cdot)\|_{TV} \leq \delta(s) V(x)
\end{equation}
for all $s \geq 0$ and $x \in \mathbb{R}^{d}$. Again, this inequality should hold deterministically, which is an $\mathcal{F}_{s}$-measurable event. For notational convenience, we define $\delta(s) = \delta(0)$ for all $s < 0$.
\label{assumption:good-approximation}
\end{assumption}

\subsubsection{Initial coupling bounds}

The following is our main technical lemma. It is \textit{not} monotone in the time $s$; this will be remedied in applications.
\begin{lemma}
 Let Assumptions \ref{assumption:geometric-ergodicity}, \ref{assumption:lyapunov-inequality}, and \ref{assumption:good-approximation} hold. There exists a constant $0 < C < \infty$ depending only on $\alpha$, $\beta$, $R$, and $\gamma$ so that for $x \in \mathbb{R}^{d}$ and triple $\{\hat{X}_t, \hat{K}_t, \mathcal{F}_t\}_{t \geq 0}$ started at $\hat{X}_0 = x$, we have 
 \begin{equation*}
\| \mathbb{P}[\hat{X}_s \in \cdot] - \pi(\cdot) \|_{TV} \leq \begin{cases}
1 & \mbox{if } s \leq C_0 \\
 C \delta(0) s & \mbox{if } s > C_0,
\end{cases}
 \end{equation*}
 where $C_0 = C \log{(\delta(0)^{-1}V(x))}$.
 \label{lem:approx-chain-bound}
\end{lemma}

\begin{proof}
Define the ``small set'' 
\begin{equation} \label{EqSmallSet}
    \mathcal{C} = \left\{y : V(y) \leq \frac{4 \beta}{\alpha}\right\}    
\end{equation}
and the associated hitting time 
\begin{equation} \label{EqSmallSetHitting}
    \tau_{\mathcal{C}} = \min{\{t : \hat{X}_t \in \mathcal{C}\}}.
\end{equation}
Denote by $T_b \geq 0$ a ``burn-in'' time whose value will be fixed toward the end of the proof. 

By the triangle inequality, for all measurable $A \subset \mathbb{R}^{d}$
\begin{eqnarray}
\vert \mathbb{P}[\hat{X}_s \in A] - \pi(A) \vert &\leq& \vert \mathbb{P}[\hat{X}_s \in A, \tau_{\mathcal{C}} \leq T_b] - \pi(A) \mathbb{P}[\tau_{\mathcal{C}} \leq T_b] \vert \nonumber  + \vert \mathbb{P}[\hat{X}_s \in A, \tau_{\mathcal{C}} > T_b] - \pi(A) \mathbb{P}[\tau_{\mathcal{C}} > T_b] \vert \nonumber \\
&\leq& \vert \mathbb{P}[\hat{X}_s \in A, \tau_{\mathcal{C}} \leq T_b] - \pi(A) \mathbb{P}[\tau_{\mathcal{C}} \leq T_b] \vert \nonumber  + \mathbb{P}[\tau_{\mathcal{C}} > T_b] \label{eq:chain-inequality}
\end{eqnarray}
To bound the first term, note that
\begin{eqnarray}
    \vert \mathbb{P}[\hat{X}_s \in A, \tau_{\mathcal{C}} \leq T_b] - \pi(A) \mathbb{P}[\tau_{\mathcal{C}} \leq T_b] \vert \nonumber  &\leq& \sup_{y \in \mathcal{C}, 0 \leq u \leq T_b}{\vert \mathbb{P}[\hat{X}_s\in A \vert \hat{X}_u = y, \tau_{\mathcal{C}}=u] - \pi(A) \vert} \nonumber \\
    &\leq& \sup_{y \in \mathcal{C}, 0 \leq u \leq T_b}{\vert \mathbb{P}[\hat{X}_s\in A \vert \hat{X}_u = y, \tau_{\mathcal{C}}=u] - K^{s-u-1}(y, A) \vert} \nonumber \\ && + \sup_{y \in \mathcal{C}, 0 \leq u \leq T_b}{\vert K^{s-u-1}(y, A) - \pi(A) \vert}.
    \label{eq:chain-inequality-first-term}
\end{eqnarray}
Assumption \ref{assumption:geometric-ergodicity} gives 
\begin{equation*}
    \sup_{y \in \mathcal{C}, 0 \leq u \leq T_b}{\left| K^{s-u-1}(y, A) - \pi(A)\right|} \leq R \gamma^{s-T_b-1}    
\end{equation*}
and Assumption \ref{assumption:good-approximation} gives 
\begin{equation*}
  \sup_{y \in \mathcal{C}, 0 \leq u \leq T_b}{\vert 
\mathbb{P}[\hat{X}_s\in A \vert \hat{X}_u = y, \tau_{\mathcal{C}}=u] - K^{s-u-1}(y, A) \vert} \leq \sup_{y \in \mathcal{C}, 0 \leq u \leq T_b}{\sum_{t=u}^{s-1} \delta(t) (K^{s-t-1} V)(y)}.
\end{equation*}
Furthermore,
\begin{equation*}
    \sup_{y \in \mathcal{C}, 0 \leq u \leq T_b}{\sum_{t=u}^{s-1} \delta(t) (K^{s-t-1} V)(y)} +  R \gamma^{s-T_b-1}  \leq \sup_{y \in \mathcal{C}}{\sum_{t=0}^{s-1} \delta(t) (K^{s-t-1} V)(y)} +  R \gamma^{s-T_b-1}.
\end{equation*}
Substituting this back into \eqref{eq:chain-inequality-first-term} and by Assumption \ref{assumption:lyapunov-inequality}, 
\begin{subequations}
\begin{eqnarray}
    \vert \mathbb{P}[\hat{X}_s \in A, \tau_{\mathcal{C}} \leq T_b] - \pi(A) \mathbb{P}[\tau_{\mathcal{C}} \leq T_b] \vert  &\leq& \sup_{y \in \mathcal{C}}{\sum_{t=0}^{s-1} \delta(t) ((1-\alpha)^{s-t-1} V(y) + \beta/\alpha)} +  R \gamma^{s-T_b-1} \\
    &\leq& \sum_{t=0}^{s-1} \delta(t) ((1-\alpha)^{s-t-1} 4 \beta/\alpha + \beta/\alpha) +  R \gamma^{s-T_b-1} \\
    &\leq& \delta(0) 5 s \beta /\alpha +  R \gamma^{s-T_b-1}.
\end{eqnarray}
\label{eq:chain-inequality-first-term-bound}
\end{subequations}

To bound the second term in \eqref{eq:chain-inequality}, recall from Assumption \ref{assumption:lyapunov-inequality} that 
\begin{eqnarray*}
    \mathbb{E}[V(\hat{X}_{t+1}) \mathbf{1}_{\tau_{\mathcal{C}} > t} \vert \mathcal{F}_t] &\leq& ((1-\alpha)V(\hat{X}_t) + \beta) \mathbf{1}_{V(\hat{X}_t)>4 \beta/\alpha} \\
    &\leq& \left(1-\frac{3\alpha}{4}\right)V(\hat{X}_t) + \left(\beta - \frac{\alpha}{4} V(\hat{X}_t)\right) \mathbf{1}_{V(\hat{X}_t)>4 \beta/\alpha} \\
    &\leq& \left(1-\frac{3\alpha}{4}\right)V(\hat{X}_t) + \left(\beta - \frac{\alpha}{4} \frac{4 \beta}{\alpha}\right) \mathbf{1}_{V(\hat{X}_t)>4 \beta/\alpha} \\
    &\leq& \left(1-\frac{3\alpha}{4}\right)V(\hat{X}_t)
\end{eqnarray*}
for all $t \geq 0$. Iterating, we find by induction on $t$ that 
\begin{equation*}
    \mathbb{E}[V(\hat{X}_{t}) \mathbf{1}_{\tau_{\mathcal{C}} > t}] \leq \left(1-\frac{3\alpha}{4}\right)^{t} V(\hat{X}_0) = \left(1-\frac{3\alpha}{4}\right)^{t} V(x).   
\end{equation*}
Thus, by Markov's inequality,
\begin{equation}
    \mathbb{P}[\tau_{\mathcal{C}}>T_b] = \mathbb{P}\left[V(\hat{X}_{\tau_{\mathcal{C}}}) \mathbf{1}_{\tau_{\mathcal{C}} > T_b}>\frac{4 \beta}{\alpha} \right] \leq \frac{\alpha}{4 \beta}\left(1-\frac{3\alpha}{4}\right)^{T_b} V(x).
    \label{eq:chain-inequality-second-term-bound}
\end{equation}
Combining \eqref{eq:chain-inequality-first-term-bound} and \eqref{eq:chain-inequality-second-term-bound}, we have shown that
\begin{equation}
\vert \mathbb{P}[\hat{X}_s \in A] - \pi(A) \vert \leq \delta(0) \frac{5 s \beta}{\alpha} +  R \gamma^{s-T_b-1} + \frac{\alpha}{4 \beta}\left(1-\frac{3\alpha}{4}\right)^{T_b} V(x).
\end{equation}
Finally, we can choose $T_b$. Set 
\begin{equation*}
    T(s) = \max{\{t : R \gamma^{s-t-1}, \frac{\alpha}{4\beta}\left(1-\frac{3\alpha}{4}\right)^{t} V(x) \leq \frac{1}{2}\delta(0)\}}    
\end{equation*}
and define $T_b = \lfloor T(s) \rfloor$ when $0 < T(s) < \infty$ and $T_b = 0$ otherwise. Define $S=\min{\{s : T(s) \in (0, \infty)\}}$. Noting that $S = \Theta(\log{(\delta(0)^{-1})}+\log{(V(x))})$ for fixed $\alpha$, $\beta$, $R$, and $\gamma$ completes the proof.
\begin{flushright}$\qed$\end{flushright}
\end{proof}

We strengthen Lemma \ref{lem:approx-chain-bound} by first observing that, if $\hat{X}_0$ satisfies $V(\hat{X}_0) \leq 4 \beta / \alpha$, then
\begin{equation}
    \mathbb{E}[V(\hat{X}_1)] \leq (1-\alpha) \mathbb{E}[ V(\hat{X}_0)]+ \beta \leq (1-\alpha) \frac{4 \beta}{\alpha} + \beta \leq \frac{4 \beta}{\alpha}.
\end{equation}
Thus, by induction, if $\mathbb{E}[V(\hat{X}_{0})] \leq 4 \beta / \alpha$, then $\mathbb{E}[V(\hat{X}_s)] \leq 4 \beta / \alpha$ for \textit{all} time $s \geq 0$. Using this (and possibly relabelling the starting time to the quantity denoted by $T_{0}(s)$), Lemma \ref{lem:approx-chain-bound} has the immediate slight strengthening:
\begin{lemma}
  Let Assumptions \ref{assumption:geometric-ergodicity}, \ref{assumption:lyapunov-inequality}, and \ref{assumption:good-approximation} hold. There exists a constant $0 < C < \infty$ depending only on $\alpha$, $\beta$, $R$, and $\gamma$ so that for all starting distributions $\mu$ on $\mathbb{R}^{d}$ with $\mu(V) \leq 4 \beta / \alpha$ and triple $\{\hat{X}_t, \hat{K}_t, \mathcal{F}_t\}_{t \geq 0}$ started at $\hat{X}_0 \sim \mu$, we have 
  \begin{equation*}
\| \mathbb{P}[\hat{X}_s \in \cdot] - \pi(\cdot) \|_{TV} \leq \begin{cases}
1, & s \leq C_0 \\
  C \delta( T_{0}(s) ) \log{(\delta(T_{0}(s))^{-1})}, & s > C_0,
\end{cases}
  \end{equation*}
where $C_0 = C \delta(0) \log{(\delta(0)^{-1})}$ and 
\begin{equation}
T_{0}(s) = s - C \delta(0) \log{(\delta(0)^{-1})}.
\label{EqT0Def}
\end{equation}
  \label{lem:approx-chain-bound-distribution}
\end{lemma}

\subsubsection{Application to bounds on mean-squared error}

Recall that $f : \mathbb{R}^{d} \to [-1,1]$ with $\pi(f) = 0$. We apply Lemma \ref{lem:approx-chain-bound-distribution} to obtain the following bound on the Monte Carlo bias:
\begin{lemma}
  (Bias estimate). Let Assumptions \ref{assumption:geometric-ergodicity}, \ref{assumption:lyapunov-inequality}, and \ref{assumption:good-approximation} hold. There exists a constant $0 < C < \infty$ depending only on $\alpha$, $\beta$, $R$, and $\gamma$ so that for all starting points $x \in \mathbb{R}^{d}$ with $V(x) \leq 4 \beta / \alpha$ and triple $\{\hat{X}_t, \hat{K}_t, \mathcal{F}_t\}_{t \geq 0}$ started at $\hat{X}_0 = x$ we have 
  \begin{equation*}
    \left| \mathbb{E}\left[ \frac{1}{T} \sum_{t=1}^{T} f(\hat{X}_t) \right] \right|  \ \leq \ \begin{cases}
    1, & T \leq C_0 \\
    \frac{C_0}{T} + \frac{C}{T} \displaystyle\sum_{s=C_0}^{T} \delta(T_{0}(s)) \log{(\delta(T_{0}(s))^{-1})} & T > C_0,
    \end{cases}
  \end{equation*}
where $C_0 = C \delta(0) \log{(\delta(0)^{-1})}$.
\label{lem:bias-estimate}
\end{lemma}
\begin{proof}
 In the notation of Lemma \ref{lem:approx-chain-bound-distribution}, we have for $T > C_0$ sufficiently large
\begin{eqnarray*} 
 \left|\mathbb{E}\left[ \frac{1}{T} \sum_{t=1}^{T} f(\hat{X}_t) \right] \right| & \leq &  \frac{1}{T} \sum_{t=1}^{C_0} |\mathbb{E}[f(\hat{X}_{t})]| + \frac{1}{T} \sum_{t=C_0}^{T} |\mathbb{E}[f(\hat{X}_{t})]| \\
&\leq& \frac{C_0}{T} + \frac{C}{T} \sum_{s=C_0}^{T} \delta(T_{0}(s)) \log(\delta(T_{0}(s))^{-1}). 
\end{eqnarray*}
\begin{flushright}$\qed$\end{flushright}
\end{proof}
We have a similar bound for the Monte Carlo variance: 
\begin{lemma}
  (Covariance estimate). Let Assumptions \ref{assumption:geometric-ergodicity}, \ref{assumption:lyapunov-inequality}, and \ref{assumption:good-approximation} hold. There exists a constant $0 < C < \infty$ depending only on $\alpha$, $\beta$, $R$, and $\gamma$ so that for all starting points $x \in \mathbb{R}^{d}$ with $V(x) \leq 4 \beta / \alpha$ and triple $\{\hat{X}_t, \hat{K}_t, \mathcal{F}_t\}_{t \geq 0}$ started at $\hat{X}_0 = x$ we have 
  \begin{equation*}
      \sqrt{\vert\mathbb{E}[f(\hat{X}_s) f(\hat{X}_t)] - \mathbb{E}[f(\hat{X}_s)] \mathbb{E}[f(\hat{X}_t)] \vert} \  \leq \ \begin{cases}
1, & m(s,t) \leq C_0 \\
  C \delta( T_{0} ) \log{(\delta(T_{0})^{-1})}, & m(s,t) > C_0,
\end{cases}
  \end{equation*}
where $m(s,t) = \min(s,t,|t-s|)$, $T_{0} = T_{0}(m(s,t))$ is as in \eqref{EqT0Def}, and $C_0 = C \delta(0) \log{(\delta(0)^{-1})}$ as before.
\label{lem:covariance-estimate}
\end{lemma}
\begin{proof}
 By the triangle inequality 
 \begin{equation*}
    \vert\mathbb{E}[f(\hat{X}_s) f(\hat{X}_t)] - \mathbb{E}[f(\hat{X}_s)] \mathbb{E}[f(\hat{X}_t)] \vert \leq \vert\mathbb{E}[f(\hat{X}_s) f(\hat{X}_t)] \vert + \vert \mathbb{E}[f(\hat{X}_s)] \mathbb{E}[f(\hat{X}_t)] \vert.
 \end{equation*}
 As above, applying Lemma \ref{lem:approx-chain-bound-distribution} completes the proof.
 \begin{flushright}$\qed$\end{flushright}
\end{proof}

The above bias and variance estimates immediate imply our main theorem on the total error of the Monte Carlo estimator: 

\begin{theorem} \label{thm_main_mce}
Let Assumptions \ref{assumption:geometric-ergodicity}, \ref{assumption:lyapunov-inequality}, and \ref{assumption:good-approximation} hold. There exists a constant $0 < C < \infty$ depending only on $\alpha$, $\beta$, $R$, and $\gamma$ so that for all starting points $x \in \mathbb{R}^{d}$ with $V(x) \leq 4 \beta / \alpha$ and triple $\{\hat{X}_t, \hat{K}_t, \mathcal{F}_t\}_{t \geq 0}$ started at $\hat{X}_0 = x$ we have 
\begin{equation*}
\mathbb{E}\left[ \left( \frac{1}{T} \sum_{t=1}^{T} f(\hat{X}_t) \right)^2 \right]  \leq \frac{2C_{0}}{T^{2}}\sum_{s=1}^{T} C(s) + \frac{3}{T} \sum_{s=1}^{T} C(s)^{2},
\end{equation*}
where $C(s) = C \delta(T_{0}(s)) \log(\delta(T_{0}(s))^{-1})$ and we write $C_{0} = C_{0}(0)$.
\label{thm:expected-error}
\end{theorem}
\begin{proof}
We calculate
\begin{eqnarray*}
\mathbb{E}\left[ \left( \frac{1}{T} \sum_{t=1}^{T} f(\hat{X}_t) \right)^2 \right]
&=& T^{-2}\left[\sum_{t=1}^{T} \mathbb{E}[f(\hat{X}_{t})^{2}] \ + \sum_{s,t \, : \, m(s,t) < C_{0}} \mathbb{E}[f(\hat{X}_{s}) f(\hat{X}_{t})]  \ + \sum_{s,t \, : \, m(s,t) \geq C_{0}} \mathbb{E}[f(\hat{X}_{s}) f(\hat{X}_{t})] \right] \\
&\leq& T^{-2} \left[\sum_{s=1}^{T} C_{0}(s) + C_{0} \sum_{s=1}^{T} C_{0}(s) + 3 T \sum_{s=1}^{T} C_{0}(s)^{2} \right] \\
&\leq& \frac{2C_{0}}{T^{2}}\sum_{s=1}^{T} C_{0}(s) + \frac{3}{T} \sum_{s=1}^{T} C_{0}(s)^{2}.
\end{eqnarray*}
\begin{flushright}$\qed$\end{flushright}
\end{proof}

\subsection{Inheriting Lyapunov conditions} \label{SubsecInheritLyap}

Observe that each step of the main ``for'' loop in Algorithm \ref{alg:la-mcmc} determines an entire transition kernel from \textit{any} starting point; denote the kernel in step $t$ by $\mathcal{K}_{t}$. Finally, let $\mathcal{F}_{t}$ be the associated filtration.

\begin{lemma} \label{LemmaLyapSimpleGen}
Let Assumptions \ref{assumption:lyapunov-inequality-simple} and \ref{assumption:good-appr} hold. Then in fact Assumption \ref{assumption:lyapunov-inequality} holds as well. 
\label{lem:lyapunov-correction}
\end{lemma}
\begin{proof}
Under Assumption \ref{assumption:good-appr}, all proposals that \textit{decrease} $V$ are \textit{more} likely to be accepted under $\hat{K}_{t}$ than under $K$, while all proposals that \textit{increase} $V$ are \textit{less} likely to be accepted under $\hat{K}_{t}$ than under $K$. Thus, for all $x$ and all $t$,
\begin{equation}
(\hat{K}_{t}V)(\hat{X}_{t}) \leq (KV)(\hat{X}_{t}) \leq (1 - \alpha) V(\hat{X}_{t}) + \beta,
\end{equation}
which completes the proof.
\begin{flushright}$\qed$\end{flushright}
\end{proof}

\subsection{Final estimates}

We combine the theoretical results in the previous sections to obtain a final estimate on the error of our algorithm. Continuing the notation as above, we have our main theoretical result: Theorem \ref{thm:convergence-rate}, whose proof we give here.

\begin{proof}
We set some notation. Define
\begin{equation}
    E(T) \equiv \mathbb{E}\left[ \left( \frac{1}{T} \sum_{t=1}^{T} f(\hat{X}_t) \right)^2 \right]. 
\end{equation}
Also define the ``burn-in" time $T_{b} = \log(T)^{2}$, and define the hitting time $\tau_{\mathcal{C}}$ as in Equation \eqref{EqSmallSetHitting}.

By Lemma \ref{LemmaLyapSimpleGen}, Assumption \ref{assumption:lyapunov-inequality} in fact holds. Note that our assumptions also immediately give Assumption \ref{assumption:good-approximation} with
\[
\delta(t) \leq 2 \gamma_{0} \sqrt{\frac{\tau_{0}}{t}}.
\]
Thus, applying Theorem \ref{thm_main_mce} in line 3 and then Assumption  \ref{assumption:lyapunov-inequality} and Markov's inequality in line 4, we have (in the notation of that theorem and assumption): 

\begin{align*}
E(T) &\leq \mathbb{E}\left[ \left( \frac{1}{T} \sum_{t=1}^{T_{b}} f(\hat{X}_t) \right)^2 \right] + \mathbb{E}\left[ \left( \frac{1}{T} \sum_{t=T_{b}+1}^{T} f(\hat{X}_t) \right)^2 \right] + 2 \mathbb{E}\left[ \frac{1}{T^{2}} \left( \sum_{t=1}^{T_{b}} f(\hat{X}_t) \right) \left( \sum_{t=T_{b}+1}^{T} f(\hat{X}_t) \right) \right]\\
&\leq \mathbb{E}\left[ \left( \frac{1}{T} \sum_{t=1}^{T_{b}} f(\hat{X}_t) \right)^2 \right] + \frac{T_{b}^{2}}{T^{2}} + \frac{2T_{b}}{T} \\
&\leq \frac{2C_{0}}{T^{2}}\sum_{s=1}^{T} C_{0}(s) + \frac{3}{T} \sum_{s=1}^{T} C_{0}(s)^{2} + \mathbb{P}[\tau_{\mathcal{C}} > T_{b}] + \frac{T_{b}^{2}}{T^{2}} + \frac{2T_{b}}{T} \\
&\leq \frac{2C_{0}}{T^{2}}\sum_{s=1}^{T} C_{0}(s) + \frac{3}{T} \sum_{s=1}^{T} C_{0}(s)^{2} + \frac{\alpha}{4 \beta} (1-\alpha)^{T_{b}} V(x) + \frac{T_{b}^{2}}{T^{2}} + \frac{2T_{b}}{T} \\
&=O\left ( \frac{1}{T^{2}} \sum_{s=1}^{T} \sqrt{\frac{\tau_{0}}{s}} \log\left (\sqrt{\frac{\tau_{0}}{s}} \right )  + \frac{1}{T} \sum_{s=1}^{T}\frac{\tau_{0}}{s} \log\left (\sqrt{\frac{\tau_{0}}{s}}\right )^{2} + \frac{\log(T)^{2}}{T} \right )\\
&= O \left ( \frac{\log(T)^{3}}{T} \right ).
\end{align*}
\begin{flushright}$\qed$\end{flushright}
\end{proof}



\end{document}